\documentclass[useAMS, usenatbib]{mn2e}
\usepackage{graphicx}
\usepackage[authoryear]{natbib}
\usepackage{times}

\makeatletter
\providecommand{\boldsymbol}[1]{\mbox{\boldmath $#1$}}

\providecommand{\tabularnewline}{\\}

\newcommand{\aap}{A\&A}
\newcommand{\aaps}{A\&AS}
\newcommand{\apj}{ApJ}
\newcommand{\apjl}{\apj}

\newcommand{\aj}{AJ}
\newcommand{\mnras}{MNRAS}
\newcommand{\apjs}{ApJS}

\newcommand{\apss}{Ap\&SS}

\newcommand{\kmps}{\mathrm{km~s^{-1}}}
\newcommand{\ion}[2]{#1$\,${\sc {#2}}}   
\newcommand{\Kelvin}{\mathrm{K}}
\newcommand{\Msun}{\mathrm{M_{\sun}}}
\newcommand{\Rsun}{\mathrm{R_{\sun}}}

\newcommand{\MsunPerYear}{\mathrm{M_{\sun}\,yr^{-1}}}

\makeatother

\begin{document}
\title[Line profile simulations]{Three-dimensional
simulations of rotationally-induced line variability from a Classical
T~Tauri star with a misaligned magnetic dipole}

\author[R. Kurosawa
et\,al.]{Ryuichi Kurosawa$^1$\thanks{E-mail:rk@physics.unlv.edu},
M.~M.~Romanova$^2$ and Tim~J.~Harries$^3$ \\ $^1$Department of Physics
and Astronomy, University of Nevada Las Vegas, Box 454002, 4505
Maryland Pkwy, Las Vegas, NV 89154-4002, USA. \\ $^2$Department of
Astronomy, Cornell University, 410~Space Sciences Building, Ithaca, NY
14853-6801, USA. \\ $^3$School of Physics, University of Exeter,
Stocker Road, Exeter EX4~4QL, UK. }

\date{Dates to be inserted}


\maketitle

\label{firstpage}

\begin{abstract}

We present three-dimensional (3-D) simulations of rotationally induced
line variability arising from complex circumstellar environment of
classical T~Tauri stars (CTTS) using the results of the 3-D
magnetohydrodynamic (MHD) simulations of Romanova et al., who
considered accretion onto a CTTS with a misaligned dipole magnetic
axis with respect to the rotational axis. The density, velocity and
temperature structures of the MHD simulations are mapped on to the
radiative transfer grid, and corresponding line source function and
the observed profiles of neutral hydrogen lines (H$\beta$, Pa$\beta$
and Br$\gamma$) are computed using the Sobolev escape probability
method. We study the dependency of line variability on inclination
angles ($i$) and magnetic axis misalignment angles ($\Theta$).  We
find the line profiles are relatively insensitive to the details of the
temperature structure of accretion funnels, but are influenced more by
the mean temperature of the flow and its geometry.  By comparing our
models with the Pa$\beta$ profiles of 42 CTTS observed by Folha \&
Emerson, we find that models with a smaller misaligngment angle
($\Theta < \sim 15^{\circ}$) are more consistent with the observations
which show that majority of Pa$\beta$ are rather symmetric around the
line centre.  For a high inclination system with a small dipole
misalignment angle ($\Theta \approx 15^{\circ}$), only one accretion
funnel (on the upper hemisphere) is visible to an observer at any 
given rotational phase. This can cause an anti-correlation of the line
equivalent width in the blue wing ($v<0$) and that in the red wing
($v>0$) over a half of a rotational period, and a positive correlation
over other half. We find a good overall
agreement of the line variability behaviour predicted by our model
and those from observations.

\end{abstract}


\section{Introduction}

\label{sec:intro}

Classical T Tauri stars (CTTS) are thought to accrue material from
their circumstellar discs via magnetospheric accretion (MA). In this
paradigm, the magnetic field of the protostar truncates the disc at a
range of radii about corotation, from where the material flows
along the field lines and onto the photosphere. The kinetic power of
the material is thermalized in shocks (e.g.~~\citealt{camenzind:1990};
\citealt{koenigl:1991}) and is observed as a blue continuum excess
(e.g.~\citealt{calvet:1998}), while the hot material within the 
funnel flows emits strongly in permitted lines
(e.g.~\citealt{alencar:2000}). Observational and theoretical aspects
of the MA paradigm were recently reviewed by 
\cite{ppv:2007}. 

There is now widespread observational support for the MA model: CTTS
are observed to have kilogauss magnetic fields that are persistent
over timescales of years (\citealt{johnskrull:1999};
\citealt{symington:2005a}; \citealt{johnskrull:2007}); the line
profiles of hydrogen are Doppler broadened to a width comparable with
the stellar escape velocity and the line profiles often show an
inverse P~Cygni profile that arises when an accretion funnel is
viewed against a hot spot (e.g. \citealt{edwards:1994};
\citealt{alencar:2000}; \citealt{folha:2001}); time-dependent line
profile studies indicate the the line profiles are modulated on the
stellar rotation period (e.g. \citealt{johns:1995};
\citealt{bouvier:2007}).

Radiative-transfer models based on the MA paradigm are broadly
successful in predicting line profile strengths and
morphologies. Initial models were based on an simple aligned-dipole
geometry for the accretion flow, but incorporated increasingly
sophisticated physics starting from a two-level atom approximation
(\citealt*{hartmann:1994}), through a full statistical equilibrium
calculation under the Sobolev approximation (SA) by
\cite*{muzerolle:1998a}, and finally including SA plus an exact
integration of the line profile (\citealt{muzerolle:2001}).

The overwhelming evidence for variability, both in the continuum
excess and the lines, of the emission from CTTS has lead us to examine
departures from axisymmetry in the accretion geometry. A crude
`curtains of accretion' model (\citealt*{symington:2005}) was able to
approximate rather well to the observed variability, providing that
the accretion curtains had to have a relatively large azimuthal
extent. A similar, tailored model for SU~Aur was also able to
reproduce some of the observed variability characteristics
(\citealt*{kurosawa:2005}).

The last few years has seen the publication of a series of papers
dealing with the magnetohydrodynamical modelling of accretion onto
CTTS (\citealt{romanova:2002}; \citealt{romanova:2003};
\citealt{romanova:2004}, hereafter R02, R03 and R04 respectively;
\citealt*{long:2007}). But how well do these MHD models agree with the
observational data? The simplest test is to attempt to model continuum
light curves from the hot spot distributions of the models, and such
simulations can reproduce the wide variety of observed
variability. In particular for models with small
magnetic misalignment angles the accretion funnels may rotate faster
or slower than the star (meaning that the hot spots are not at fixed
on the stellar surface), this naturally leads to the quasi-periodic
variability that is often observed (R04). Useful though such
comparisons are, the line profiles themselves encode much more
detailed information on the kinematics and geometry of the flow. The
aim of this paper is to make a quantitative comparison between the MHD
models and spectroscopic observations by computing line profiles based
on the density and velocity structure of the MHD calculations.

Here we concentrate on three hydrogen transitions (H$\beta$,
Pa$\beta$, and Br$\gamma$). The H$\alpha$ profile, although the
strongest and most widely observed optical line, is usually
contaminated by outflow emission and absorption
(e.g. \citealt*{reipurth:1996}) and cannot be modelled by
MA alone, instead requiring a hybrid code
incorporating both accretion and outflow
(e.g. \citealt*{kurosawa:2006}). The H$\beta$ line is a better proxy
for accretion, and observationally the near-IR lines show a high
frequency of inverse P~Cygni morphology, indicating that they too are
better probes of the accretion geometry (\citealt{folha:2001}).

In the following Section we describe the MHD model and the
radiative-transfer model, and we give the results of our profile
calculations in Section~3. We discuss our results in comparison with
both earlier radiative-transfer models and observations in Section~4,
and our conclusions are presented in Section~5.

\section{Models}
\label{sec:Models}

\begin{figure}

\begin{center}\includegraphics[clip,scale=0.48]{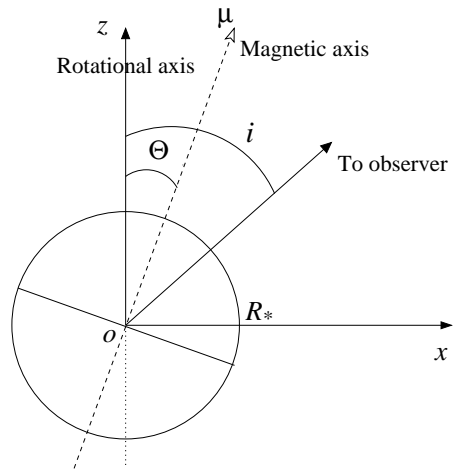}\par\end{center}

\caption{Basic model configuration. A star with its radius $R_{*}$ is
located at the origin ($O$) of cartesian coordinate system ($x$, $y$,
$z)$. The $y$-axis is into the paper. The rotational axis of the star
coincides with $z$-axis.  Its magnetic axis (with the magnetic moment
$\mu$) is inclined from the rotational axis by $\Theta$, causing
precession of magnetic axis as the star rotates. This angle will be
referred to as the misalignment angle. The inclination angle $i$ is
defined as the angle, measured from $z$-axis, to an observer located
at infinity.}

\label{fig:config}

\end{figure}

The basic model configuration of the central star is shown
Fig.~\ref{fig:config}. Two important parameters in our models, the
misalignment angle $\Theta$ and the inclination angle $i$ are also
shown in the figure. The former is defined as the angle between the
rotational axis of the star (the $z$-axis) and the magnetic axis (with
the dipole moment $\mu$), while the latter is defined as the angle between the
rotational axis and the direction to an observer.  The rotational
axes of the disc and the star coincide. In the following,
we will describe our MHD models, radiative transfer models, assumed
temperature structure and the sources of continuum radiation.

\subsection{MHD models}
\label{sub:MHD_models}

\begin{figure}
  \includegraphics[scale=0.4, angle=0]{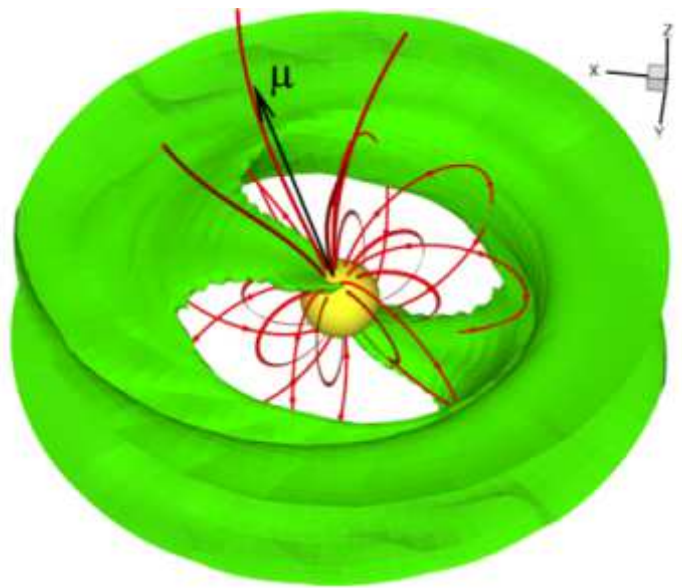}
  \caption{Example of three-dimensional simulations of
           MA flow for the magnetic axis
           misalignment angle $\Theta=15^\circ$. The background
	   shows the iso-surface for one of density levels, red lines show
	   sample magnetic field lines, and the black arrow shows the direction
	   of the magnetic moment of the star. The rotational axis
           coincides with $z$-axis.}
\label{fig:mhd-3d-density} 
\end{figure}
\begin{figure*}
 \begin{center}
  \includegraphics[angle=0,width=17cm]{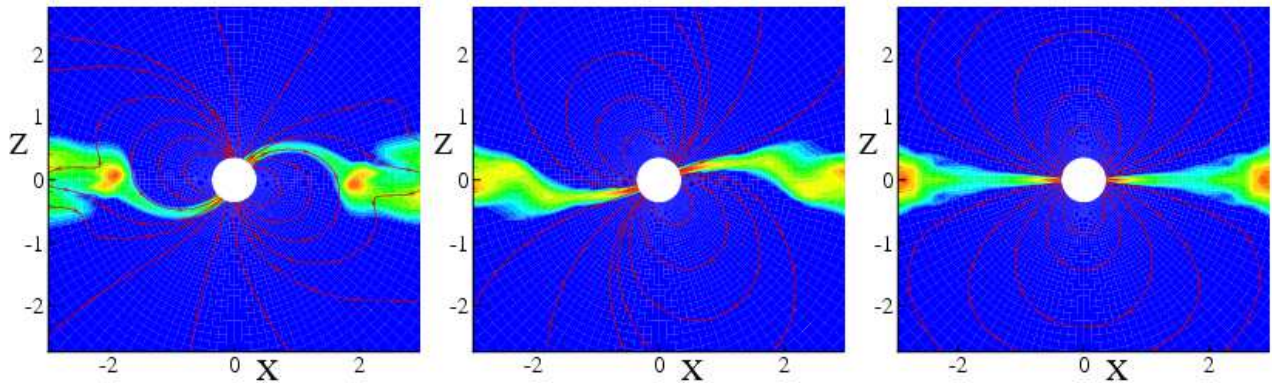}
  \caption{Slices of density distribution (background) and sample
    magnetic field lines for the simulations with the magnetic axis
    misalignment angle $\Theta=15^{\circ}$ (left), $60^{\circ}$
    (centre) and $90^{\circ}$ (right). The $X$-$Z$ plane of slices
    coincides with the plane defined by the rotational axis and the
    magnetic axis (c.f.~Fig~{\ref{fig:config}}).  The red colour
    corresponds to the maximum of the density ($\rho=2\,\rho_{0}$)
    while the dark blue colour to the minimum density
    ($\rho=0.003\,\rho_{0}$) where $\rho_{0}=4.9 \times
    10^{-12}\,\mathrm{g\,cm^{-3}}$. The units of $X$ and $Z$ dimensions
    are in $R_{0}=3.6\times 10^{11}\,\mathrm{cm}$. The accretion on to
    the surface occurs in two streams, and  the latitudinal location where
    the gas impacts on the stellar surface decreases as $\Theta$
    increases. These three density distributions (along with
    corresponding temperatures and velocities) will be used in in the
    subsequent radiative transfer calculations. }
\label{fig:density-zx} 
\end{center}
\end{figure*}

Three-dimensional MHD code and model used in this paper were
developed and described earlier in Koldoba et al. (2002), R03, and
R04. Here we briefly discuss the main aspects of the model and
also describe new simulations runs.

We investigate matter flow around a rotating star with a
misaligned dipole magnetic field. A star is surrounded with a
dense cold accretion disk and a hot low-density corona above and
below it. To calculate matter flow we solve a full system of
magnetohydrodynamic equations (in three dimensions) using a
Godunov-type numerical code (see Koldoba et al. 2002; R03 and
R04). Equations are written in the coordinate system rotating with
the star. A viscosity term has been added to the code with viscosity
coefficient proportional to the $\alpha-$ parameter.

The boundary conditions used here are similar to those in R03 and
R04. At the stellar surface, `free' boundary conditions to the
density and pressure are used. A star is treated as a perfect
conductor so that the normal component of the magnetic field does
not vary in time.  There is however a `free' condition to the
azimuthal component of the magnetic field: ${\partial(R
B_\phi)}/{\partial R}=0$ so that magnetic field lines have a
`freedom' to bend near the stellar surface. In the reference frame
rotating with the star the flow velocity is adjusted such that to
be parallel to the magnetic field $\boldsymbol{B}$ at $R=R_*$
which corresponds to a frozen-in condition. Matter falls to the
surface of the star supersonically and most of its energy is expected
to be radiated in the shock wave close to the surface of the
star (e.g.~\citealt{camenzind:1990}; \citealt{koenigl:1991};
\citealt{calvet:1998}). Evolution of the radiative 
shock wave above the surface of CTTSs has been considered in
detail by \citet{ustyugova:2006}. In this paper we suggest that
most of kinetic energy of the flow is converted to radiation  (see
 R04 and Section~\ref{sec:continuum_sources}). At the
outer boundary $R=R_{\rm max}$, free boundary conditions are taken
for all variables.
				   
Numerical simulations have shown that the inner regions of the
disk are disrupted by the magnetosphere of the star and matter
flows to a star in two high-density funnels streams under some
situations.
Fig.~\ref{fig:mhd-3d-density} shows an example of such two-stream
accretion for a system with the misalignment angle
$\Theta=15^{\circ}$ (see also R04; \citealt{kulkarni:2005}). The
important parameters are the initial densities in the disk
$\rho_d$ and corona $\rho_c$, and the initial temperatures in the
disk $T_d$ and corona $T_c$. These values are determined at the
fiducial point at the boundary between the disk and corona (at the
inner radius of the disk). We took parameters $\rho_c=0.01
\rho_d$, $T_d=0.01 T_c$ and obtained supersonic funnel streams
since the sound speed is relatively low, $c_s\approx 0.1 v_K$.

The funnel flow converges towards the star and temperature
increases due to adiabatic heating. In reality, the temperature
may decrease due to radiative cooling.  To mimic the effect of the
radiative cooling we performed most of these new runs at smaller
adiabatic index $\gamma=1.1$.  This is the main difference of the new
runs compared to R04 runs. In addition, we increased the magnetic
field of the star by a factor of two compared to R04. This led to larger
magnetospheric gaps compared to R04.

We have adopted the following MHD input parameters from a typical
CTTS. The mass and the radius of the star are assumed to be
$M_*=0.8\, \Msun$ and $R_*=1.8\,\Rsun$ respectively. The magnetic
field at the surface of the star 
(at the equator) is assumed to be $B_*=4\times 10^3~{\rm G}$. 
The size of the simulation region corresponds to
$R_{\rm max}= 40 R_* = 0.34~{\rm AU}$. 
The unit of time used in the model is $P=P_0=1.38$~d which corresponds
to a period of Keplerian rotation at $R=R_0=3.6\times 10^{11}\,{\rm cm}$
which is the unit distance used in Fig.~\ref{fig:density-zx}. 
The star rotates with period $P_*=3.9~{\rm d}$ which
corresponds to many CTTSs (e.g.~\citealt{herbst:2002}). 
The accretion disc is stopped by the magnetosphere at the distance
$R_t\approx 1.6 R_0 = 5.8\times 10^{11}\,{\rm cm}$. This distance
is slightly below the corotation radius of the star $R_{\mathrm{cor}}=2
R_0=7.2\times 10^{11} \, {\rm cm}$. This situation approximately
corresponds to the rotational equilibrium state in which a star
does not gain or lose angular momentum on average (see \citealt{long:2005}).
We consider the case with typical value of corona temperature and the 
disc density are set $T_c=4.5\times 10^6~{\rm K}$ and
$\rho_0=4.9\times 10^{-12}~{\rm g\,cm^{-3}}$
(e.g.~\citealt{hartmann:1998}) respectively.

R03 and R04 showed that the geometry of the accretion stream
strongly depends on $\Theta$; hence, we consider cases with a wide
range of $\Theta$ angles, specifically $\Theta=15^{\circ}$,
$60^{\circ}$ and $90^\circ$. Fig.~\ref{fig:density-zx} shows the
density slices of the 3-D simulations on the planes defined by the
rotation axes and the misaligned magnetic axes.  The figure shows
the accretion occurs in two streams (as mentioned earlier), and
the flows encounter the stellar surface near the magnetic poles
(see R04 for their relative locations). We choose the time slices
of these MHD simulations at which the flows are semi-steady and
has a similar accretion rate ($\dot{M}_{\mathrm{acc}} \approx
2.0\times 10^{-8}\,\MsunPerYear$) for a comparison and for the
radiative transfer models presented in the following sections.
This mass-accretion rate chosen here is very similar to that of a
typical CTTS seen in the observations
(e.g.~\citealt{gullbring:1998}; \citealt{calvet:2004}).

\subsection{Radiative transfer model}
\label{sub:RT_model}

The radiative transfer code {\sc TORUS} (\citealt{harries:2000};
\citealt{kurosawa:2004}; \citealt{kurosawa:2005}:
\citealt{symington:2005}: \citealt{kurosawa:2006}) was extended to
incorporate the density, velocity and gas temperature structures from
the 3-D MHD simulations of R04 mentioned above. The
radiative transfer code uses the three-dimensional (3D) adaptive
mesh refinement (AMR) grid, and it allows us an accurate mapping of
the original MHD simulation data onto the radiative transfer grid.
Although it is possible, we do not explore the line variability due to
the time-dependent nature of the 
accretion in this paper. The aspect we investigate here is
the variability due to the change in the viewing angles (of an
observer) due to the rotational motion of a star and its
magnetosphere.  For this reason, we select outputs of MHD simulation
which have (relatively) quiet stage, i.e. we choose the time stage of
simulations which reached a (semi-) steady state.

We emphasise that the variability associated with the time-dependent
nature of the flow (e.g.~due to instabilities) is certainly worth
pursuing in the future since it provides us an opportunity to study
the kinematics of the accretion flow itself; hence, it would provide
us an additional constraint on the geometry of the magnetosphere
around CTTS. This is beyond the scope of this paper, but should be
considered in a future work.

 The basic steps for computing the line variability are as follows:
(1) mapping of the MHD simulation output onto the radiative transfer
grid, (2) the source function ($S_{\nu}$) calculation and (3) the
observed flux/profile calculation as a function of rotational
phase. In the second step, we use the method of \citet{klein:1978}
(see also \citealt{rybicki:1978}; \citealt{hartmann:1994}) in which
the Sobolev approximation method is applied. The population of the
bound states of hydrogen are assumed to be in statistical equilibrium,
and the gas to be in radiative equilibrium.  Our hydrogen atom model
consists of 14 bound states and a continuum.  Readers are referred  to
\citet{harries:2000} for details.

Monte Carlo radiative transfer (e.g. \citealt{hillier:1991}), under
the Sobolev approximation, is valid when (1) a large velocity gradient
is present in the gas flow, and (2) the intrinsic line width is
negligible compared to 
the Doppler broadening due to the bulk (macroscopic) motion of gas.
In our earlier models (\citealt{harries:2000}; \citealt{symington:2005}), this
method was adopted since these conditions are satisfied. However, as
noted and demonstrated by \citet{muzerolle:2001}, even with a moderate
mass-accretion rate ($10^{-7}~\MsunPerYear$), Stark broadening becomes
important in the optically thick H$\alpha$ line. In addition,
H$\alpha$ is most likely affected by the wind absorption and emission
components (e.g.~\citealt{edwards:1994}; \citealt{reipurth:1996};
\citealt{kurosawa:2006}), but the original MHD simulations (R03; R04)
 do not contain outflow/wind
components. For these reasons we avoid modelling H$\alpha$, and
concentrate on lines less likely to be affected by wind and Stark
broadening, i.e.~Pa$\beta$ (mainly), Br$\gamma$, and H$\beta$ in this
paper.

When importing the MHD simulations results (Fig.~\ref{fig:density-zx})
to the radiative transfer calculations, we have introduced a cut-off
radius ($r_{\mathrm{max}} = 7.0\times10^{11}$~cm $= 5.5\,R_{*}$)
although the radial range of the MHD simulations extend much larger
than this value.  The cut-off radius corresponds $X \approx 2.0$ in
Fig.~\ref{fig:density-zx}.  This indicates that the puffed-up density
structures in the disc seen in the MHD simulations are not included in
the radiative transfer models, and the problems are rather focused on
the accretion funnel parts. This is done to avoid the complications of
adding dust opacity and finding dust temperature, and to avoid very
high density and low temperature regions in which the source function
calculation may have some difficulty.  Beyond this cut-off radius,
we simply inserted a `optically thick and geometrically thin disc'
which is essentially a geometrical disc with no thickness and with
infinitely large opacity through which no radiation can
penetrate. This type of disc was adopted to imitate the obscuration of
the accretion funnels and stellar surface by a optically thick disk.
Although we understand the importance of including the puffed-up
regions of the disc and the dust for the line variability problems in
very high inclination system, currently our model are not be able
handle the regions correctly.

\subsection{Temperature structure}

\label{sec:temp_structure}

The temperature structure of the magnetosphere used by
\citet{hartmann:1994} is computed by assuming a volumetric heating
rate which is proportional to $r^{-3}$, by solving the energy balance
of the radiative cooling rate (see Table~1 in \citealt*{hartmann:1982})
and the heating rate.  However, in this formulation, the normalisation
is arbitrary, and it has to be determined from the multiple line
fitting. On the other hand, \citet{martin:1996} presented a
self-consistent determination of the thermal structure of the
inflowing gas, along the same dipole magnetic field geometry as in
\citet{hartmann:1982}, by solving the heat equation coupled to the
rate equations for hydrogen. He found that main heat source is
adiabatic compression due to the converging nature of the flow, and
the major contributors to the cooling process are bremsstrahlung
radiation and line emission from \ion{Ca}{ii} and \ion{Mg}{ii}
ions. \citet*{muzerolle:1998} found that the line profile models
computed according to the temperature structure of \citet{martin:1996}
do not agree with observations, unlike profiles based on the (less
self-consistent) \citet{hartmann:1994} temperature distribution.  It
is clear that the temperature structure of the magnetosphere is still
a large source of uncertainty in the accretion model, and this issue
should certainly be investigated more carefully in the future.

In this paper we simply consider following three cases of the
temperature structure of the flow: (1)~Hartmann-like cooling/heating
case, (2)~adiabatic cooling/heating only case (directly from the MHD
calculation), and (3) isothermal case. In case of (2), we find that
the gas temperature from the MHD is in general too high (no radiative
cooling); hence, we introduce a scaling factor $s$ which is multiplied
with the original temperature ($T_{\mathrm{MHD}}$) of the MHD
simulations, i.e.~ $T=s\, T_{\mathrm{MHD}}$.  This is somewhat similar
to an arbitrary normalisation constant introduced in (1). Note in all
the models presented in this paper, $s=1.67\times10^{-2}$ and the MHD
models results with the adiabatic index $\gamma=1.1$
(c.f.~R04). Basic results of the dependency of
line profiles will be presented in
Section~\ref{sec:result_temp_structure}.  In the following sections,
(1) and (2) will be refer to as the HCH and ACH temperature
structures, respectively.

\subsection{The continuum sources}

\label{sec:continuum_sources}

We adopt stellar parameters of a typical classical T~Tauri star used
by R04 for the central continuum source to be consistent with their
MHD simulations, i.e.~ its stellar radius $R_{*}=1.8\,\Rsun$ and its
mass $M_{*}=0.8\,\Msun$ (see Section 2.1). Consequently, we adopt the
effective temperature of the photosphere $T_{ph}=4000\,\Kelvin$ and the
surface gravity $\log g_{*}=3.5$ (cgs), and use the model atmosphere
of \citet{kurucz:1979} as the photospheric contribution to the
continuum flux. The parameters are summarised in
Table~\ref{tab:stellar_parameters}.

Additional continuum sources to be included are the hot spots formed
by the infalling gas along the magnetic field on to the stellar
surface.  As the gas approaches the surface, it decelerates in a
strong shock, and is heated to $\sim10^{6}\,\Kelvin$. The X-ray
radiation produced in the shock will be absorbed by the gas locally,
and re-emitted as optical and UV light (\citealt{calvet:1998};
\citealt{gullbring:2000}) -- forming the high temperature regions on
the stellar surface with which the magnetic field intersects. While
\citet{muzerolle:2001}, \citet{symington:2005} and
\citet{kurosawa:2006} used a single temperature model for hot spots
assuming the free-falling kinetic energy is thermalized in the
radiating layer, and is re-emitted as blackbody radiation, we 
adopt the multi-temperature hot spot model of R04 in
which the temperature of the hot spots is also determined by
conversion of kinetic energy plus internal energy of infalling gas to
radiation energy (as a blackbody). They considered the position
dependent matter flux crossing the inner boundary; hence, achieving a
position dependent temperature of the hot spot, which can be written
as
\begin{equation}
  T_{\mathrm{hs}}=\left\{
    \frac{\rho\,\left|v_{r}\right|}{\sigma}\left(\frac{1}{2}v^{2}+w\right)^{2}\right\}^{1/4}
  \label{eq:hot_spot_romanova} 
\end{equation}
 where $p$, $\rho$, $v_{r}$ and $v$ are the pressure, the density, the radial component
of velocity and the speed of gas/plasma, respectively. Further more,
$\sigma$ is the Stefan-Boltzmann constant, and $w$ is the specific enthalpy of
the gas: $w=\gamma\left(p/\rho\right)\left(\gamma-1\right)$. 
A typical temperature (area-weighted mean) of the hotspots is around 8000~K in our models.

We compare this temperature $T_{\mathrm{hs}}$ with the effective
temperature of photosphere $T_{\mathrm{ph}}$ to determine the shape of
the hot spots. If $T_{\mathrm{hs}}>T_{\mathrm{ph}}$, then the location
on the stellar surface is flagged as hot. For the hot surface, the
total continuum flux is the sum of the blackbody radiation with
$T_{\mathrm{hs}}$ and the flux from the model photosphere mentioned
above. The contribution from the inflow gas is ignored when
$T_{\mathrm{hs}}<T_{\mathrm{ph}}$.

\section{Results}
\label{sec:results}

Using Pa$\beta$ as an example, the general characteristics of line
variability computed by the radiative transfer model will be
presented in this section.  We present models with 5 different
combinations of the misalignment angle $\Theta$ and inclination angle
$i$, as summarised in Table~\ref{tab:model_parameters}.  First, we
present the continuum variability of our models. Second, we briefly
discuss the dependency of the models on the temperature structures
(c.f.~Section~\ref{sec:temp_structure}).  Third, we discuss the
dependency on other main input parameters ($i$ and $\Theta$).  Fourth,
we compare line profiles for different transitions (H$\beta$,
Pa$\beta$ and Br$\gamma$). Finally, we will present line equivalent
widths computed as a function of rotation phase.  Unless specified
otherwise, we adopt the stellar parameters in
Table~\ref{tab:stellar_parameters} for a central star. The
mass-accretion rate ($\dot{M}_{\mathrm{acc}}$) and the inner radius
($R_{\mathrm{d}}$) of the accretion disc used in our models are also
shown in the same table.

\subsection{Continuum Variability}
\label{sub:Continuum-Variability}

\begin{figure}

\begin{center}

\includegraphics[clip,width=0.45\textwidth]{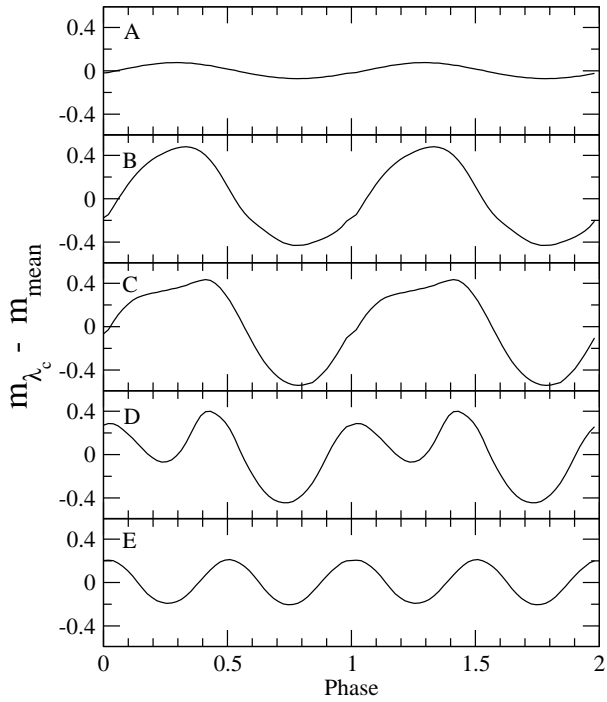}\par
				   
\end{center}

\caption{Continuum light curves for Models A to E (from top to bottom)
at the wavelength $\lambda_{c}=4800$~\AA\, (near H$\beta$). The
vertical axis is scaled in pseudo-magnitude which is defined as
$m_{\lambda_{c}}=-\log_{2.51}F_{\lambda_{c}}+m_{0}$ where $F_{\lambda_{c}}$ and
$m_{0}$ are the flux in cgs and the offset magnitude respectively. To
avoid defining the arbitrary offset value, we have subtracted the
mean magnitude $m_{\mathrm{mean}}$ (over entire rotational phases)
from $m_{\lambda_{c}}$. }

\label{fig:cont_var}

\end{figure}

Since the line variability is closely related with the continuum
variability, we present the light curves predicated by the models
before we present the line variability results. R04
also presented the light curves of the 3-D MHD simulations which are
basically identical to ours. Their light curves were computed by using
frequency-integrated flux and do not contain an optically thick disc
which would obscure the stellar photosphere and accretion hot spots at
high inclination angles. On the other hand, our model computes the
light curves at a given wavelength, and contains an optically-thick
and geometrically thin disc. Fig.~\ref{fig:cont_var} shows the light
curves computed at $\lambda_{c}=4800$~\AA\,for all the models listed
in Table~\ref{tab:model_parameters}.

Comparing the light curves from Models A, B and C which use the same
MHD model ($\Theta=15^{\circ}$ case), one can see the dependency on
the inclination angle. Note that the hot spots are located about
$30^{\circ}$ from the poles of the rotational axis for these models 
(c.f.~Figure~\ref{fig:summary_models_ABC}; also see Fig.~7 of
R04). The 
amplitude of the light curve oscillations increases as $i$ increases up
to $i\approx60^{\circ}$ (Model~B). Although the models with $i$
between $0^{o}$ and $60^{o}$ are
not shown here, the same trend is observed in the additional models we
have run. A similar trend was also observed by
R04 --- see their Fig.~10. As $i$ becomes greater than
$\sim60^{\circ}$, the peak amplitude does not change greatly since a
part of the second hot spot on the lower hemisphere becomes visible at
higher inclinations.

The visibility of the second hot spots affects the shape of the light
curve. In Model A and B, only one hot spot (on the upper hemisphere)
is visible during the whole or parts of rotational phase; hence, their
light curves exhibit a single peak in one rotational phase. On the
other hand, two hot spots are visible in Models~C, D and E
(c.f.~Figures~\ref{fig:summary_models_ABC} and
\ref{fig:summary_models_DE}). As a consequence, their light curves
have two peaks in one rotational phase. Note that the Model C does not
show a clear second peak, but the shape of the light curve is heavily
affected by the presence of the second spot.

For the models with a fixed inclination angle but with different
misalignment angles (Models~B, D and E), the oscillation amplitudes
are similar to each other except for Model~E that has a large
misalignment angle ($90^{o}$).  The shape of the hot spots in Model~E is extremely 
elongated, and has almost a belt like shape located near the equatorial
plane (c.f.~Figure~\ref{fig:summary_models_DE}; also see Fig.~2 of
R04). The gap between one edge of the {}`belt' like structure to
the next is ($\sim40^{\circ}$), causing a smaller peak-amplitude of
oscillations. 

The amplitude of the light curves ($\Delta m \approx 0.4$) of
Models~B, C and D are comparable to that of AA~Tau observed in V-band
($\Delta V \approx 0.5$) by \citet{bouvier:2007}. This suggests that
the continuum sources (Section~\ref{sec:continuum_sources}) used in
our models are quite reasonable.

\subsection{Dependency on temperatures structures}
\label{sec:result_temp_structure}

We examine how the different assumed temperature structures introduced
in Section~\ref{sec:temp_structure}  affect the line profiles.
As a demonstration, we compute Pa$\beta$ from a system viewed at
$i=60^{\circ}$ and at the rotational phase $t=0.75$, with a
misalignment angle $\Theta=15^{\circ}$ (Model~B in
Table~\ref{tab:model_parameters}; see also
Figs.~\ref{fig:mhd-3d-density} and \ref{fig:density-zx}).  

In particular, we will compare the profiles computed with the HCH, ACH
temperature structures (c.f.~Section~\ref{sec:temp_structure}), and an
isothermal case ($T_{\mathrm{iso}}=8000\,\Kelvin$).  We choose
normalisations/scalings of the temperature structures in HCH and ACH
models such that their density-weighted mean temperatures
($T_{\mathrm{mean}}$) of the gas are comparable to that of the
isothermal case ($\sim8000\,\Kelvin$).

\begin{figure}

\begin{center}
 \includegraphics[clip,scale=0.58]{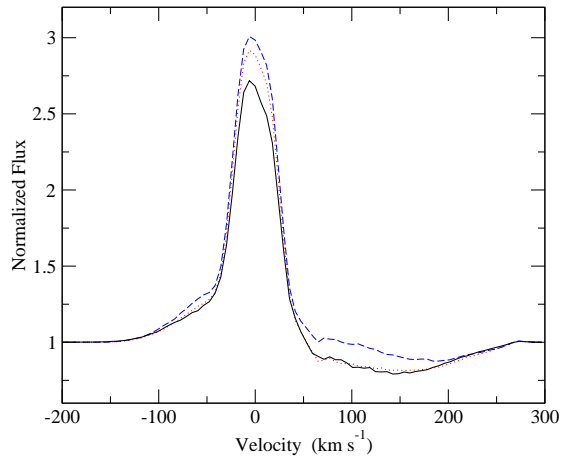}
\end{center}

\caption{Comparison of Pa$\beta$ with three different temperature
structures of the accretion stream with $\Theta=15^{\circ}$ and 
$i=60^{\circ}$ (Model~B) at the rotational phase $t=0.75$. Three
cases considered here are: (1)~the ACH model: the temperature structure from
R04 (solid), (2)~the HCH model:
temperature structure from \citet{hartmann:1994} (dotted), and (3)~an isothermal
($T=8000~\Kelvin$) case (dashed). The normalisation of the temperature
structure scalings are chosen such that the mass-weighted mean
temperature ($T_{\mathrm{mean}}$) of the flow is comparable to that of
the isothermal case, i.e. $T_{\mathrm{mean}}\approx8000~\Kelvin$. 
Overall shapes of the profiles are very similar to each other, but the
absorption in the red wing is slightly shallower for the isothermal
case compared to the other two cases. The difference in the
temperature structure in the accretion funnels does not make a
significant difference in the profile shape as long as the mean
temperature of the flows is similar. }

\label{fig:PaB_temp_structure}

\end{figure}

\begin{figure*}

\begin{center}
\includegraphics[clip,width=1.2\textwidth,angle=270]{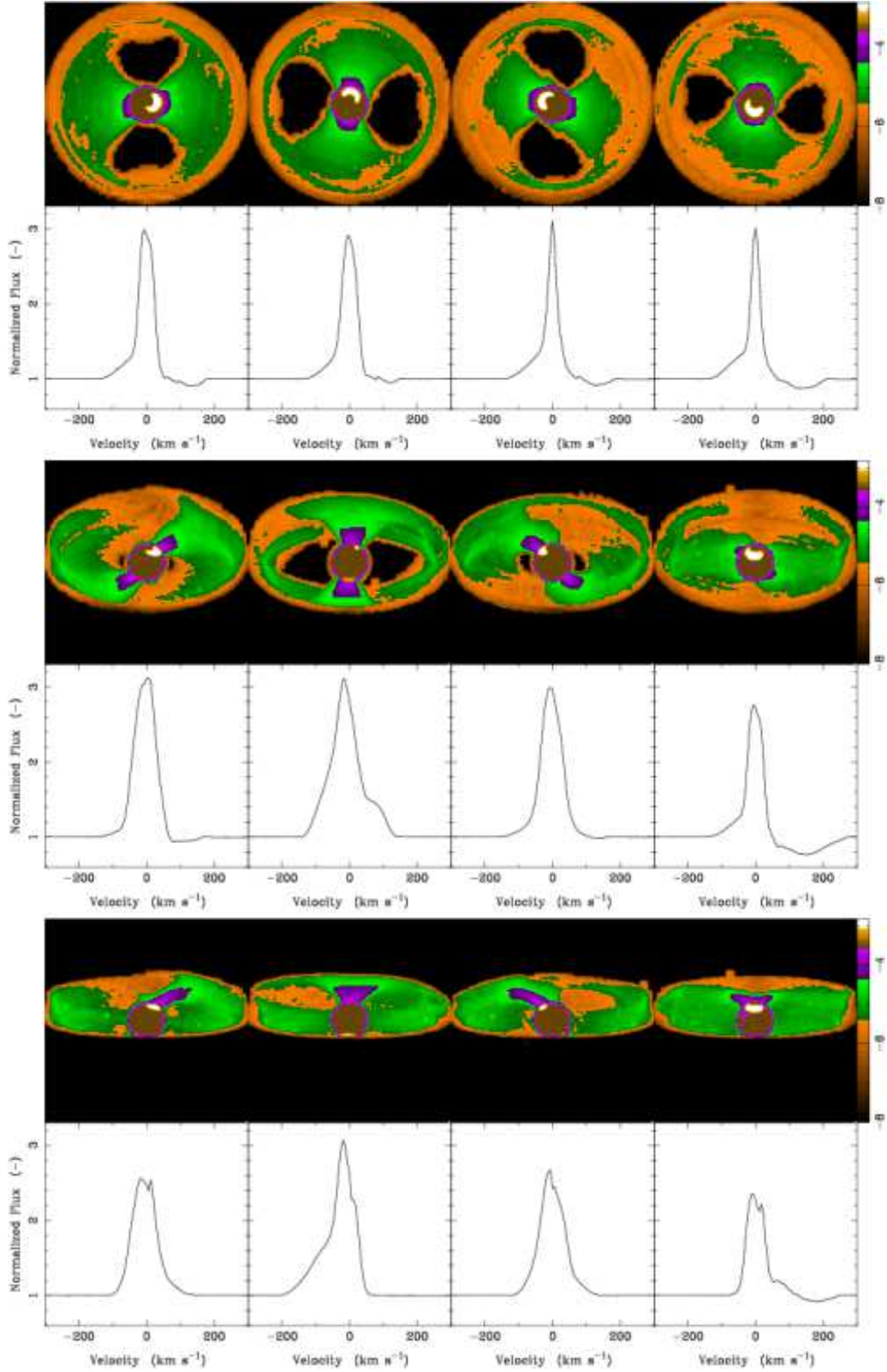}
\end{center}

\caption{Pa$\beta$ model intensity maps and the corresponding profiles computed
at rotational phases $t=0.0$, $0.25$, $0.5$ and $0.75$ (from left
to right) and for inclination angles $i=10^{\circ}$, $60^{\circ}$,
and $80^{\circ}$ (from top to bottom). The misalignment angle of the
magnetic axis is fixed at $\Theta=15^{\circ}$ for all the models shown
here. The intensity is shown in logarithmic scale with an arbitrary
units. The physical dimension of the images are $1.4\times10^{12}$~cm
( $\sim 11\,R_{*}$) in both horizontal and vertical directions.
 The top, middle and bottom correspond to Model~A, B and C
in Table~\ref{tab:model_parameters} respectively.}

\label{fig:summary_models_ABC}

\end{figure*}

\begin{figure*}

\begin{center}
\includegraphics[clip,width=0.80\textwidth,angle=270]{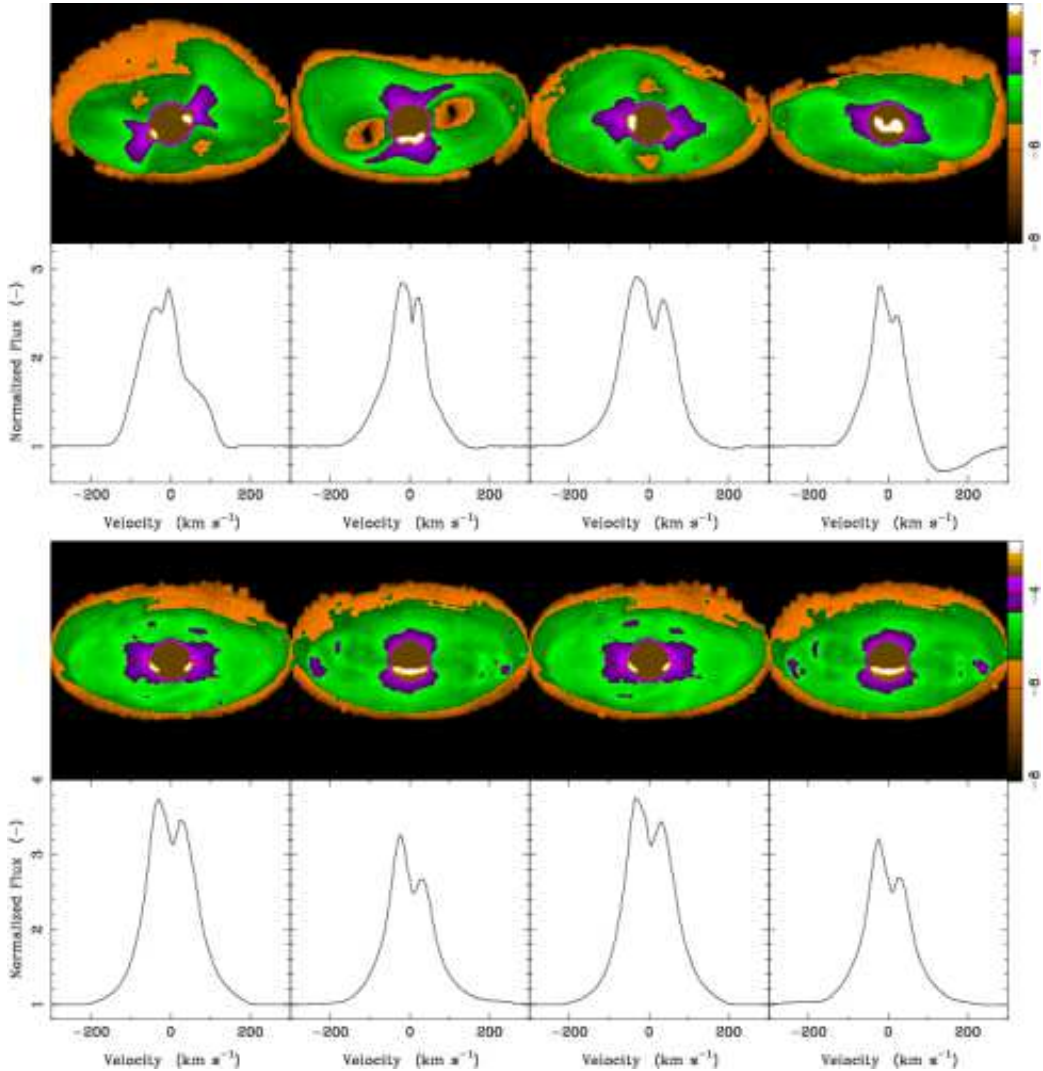}
\end{center}

\caption{Same as in Fig.~\ref{fig:summary_models_ABC}, but for Models~D
(upper panels) and E (lower panels) (c.f.~Table~\ref{tab:model_parameters}). }

\label{fig:summary_models_DE}

\end{figure*}

The results are shown in Fig.~\ref{fig:PaB_temp_structure}.  The
profiles computed with the HCH and ACH are very similar to each other -- 
they have similar flux in both wings, but the core flux of the HCH
model is slightly larger than that of the ACH model.  On the other
hand, the isothermal model produces an inverse P-Cygni (IPC) profile
with a shallower absorption wing. The difference is mainly caused by
the warmer gas stream present in the outer part of the magnetosphere
(where absorption occurs) compared to that of the HCH and ACH models.
The core flux of the isothermal case is very similar to that of the
HCH model. Overall, the profile shapes from all three temperature
structures are very similar to each other.  From this exercise, we
find that the main physical conditions which determines the profile
shapes,\, \em for a fixed mean temperature condition\em, are the
velocity field and the geometry of the funnel flows. In other words,
as long as the mean temperature is similar, the difference in the
temperature structure along the stream does not make a significant
difference in line profiles at least at the temperature used here
(8000~K).  Since we find no large difference between the profiles from
the HCH and ACH temperature structures, we adopt the latter
(with $T_{\mathrm{mean}}=8000\Kelvin$) in the models presented in the
following sections unless specified otherwise.
				   
\begin{table}

\begin{center}\begin{tabular}{ccccc}
\hline 
$M_{*}$&
$R_{*}$&
$T_{\mathrm{eff}}$&
$\dot{M}$&
$R_{\mathrm{d}}$\tabularnewline
$\left(\Msun\right)$&
$\left(\Rsun\right)$&
$\left(\Kelvin\right)$&
$\left(\MsunPerYear\right)$&
$\left(\Rsun\right)$\tabularnewline
\hline 
$0.8$&
$1.8$&
$4000$&
$\sim 2\times10^{-8}$&
$\sim 8.6$\tabularnewline
\hline
\end{tabular}\par\end{center}

\caption{Reference Model Parameters}

\label{tab:stellar_parameters}

\end{table}



\begin{table}

\begin{center}

\begin{tabular}{cccccc}
\hline 
Model &
A&
B&
C&
D&
E\tabularnewline
\hline 
Dipole offset, $\Theta$&
$15^{\circ}$&
$15^{\circ}$&
$15^{\circ}$&
$60^{\circ}$&
$90^{\circ}$\tabularnewline
Inclination, $i$&
$10^{\circ}$&
$60^{\circ}$&
$80^{\circ}$&
$60^{\circ}$&
$60^{\circ}$\tabularnewline
\hline
\end{tabular}

\end{center}

\caption{A summary of the main model parameters}

\label{tab:model_parameters}

\end{table}



\begin{figure*}

\begin{center}\begin{tabular}{ccc}
\includegraphics[clip,scale=0.3]{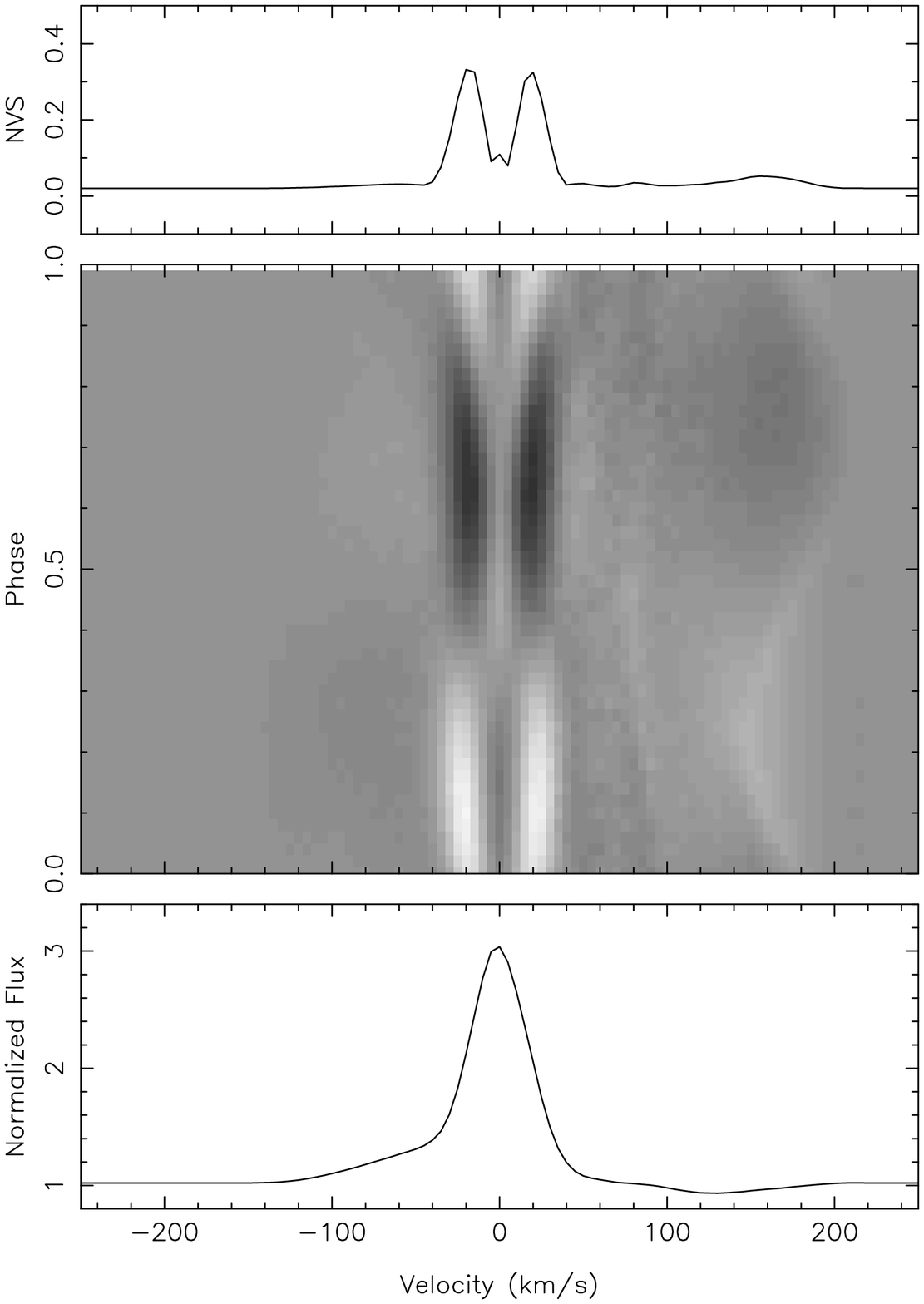}&
\includegraphics[clip,scale=0.3]{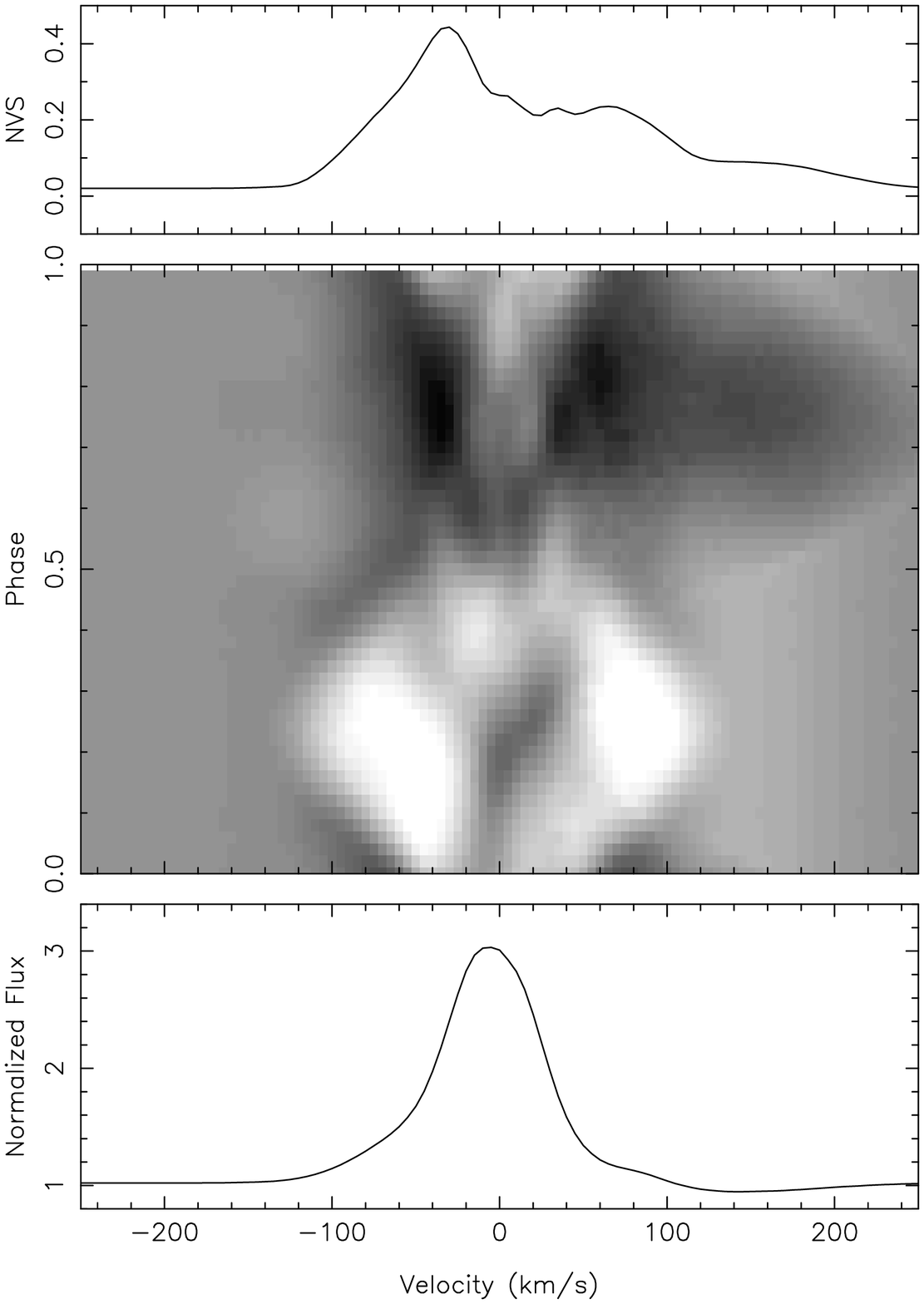}&
\includegraphics[clip,scale=0.3]{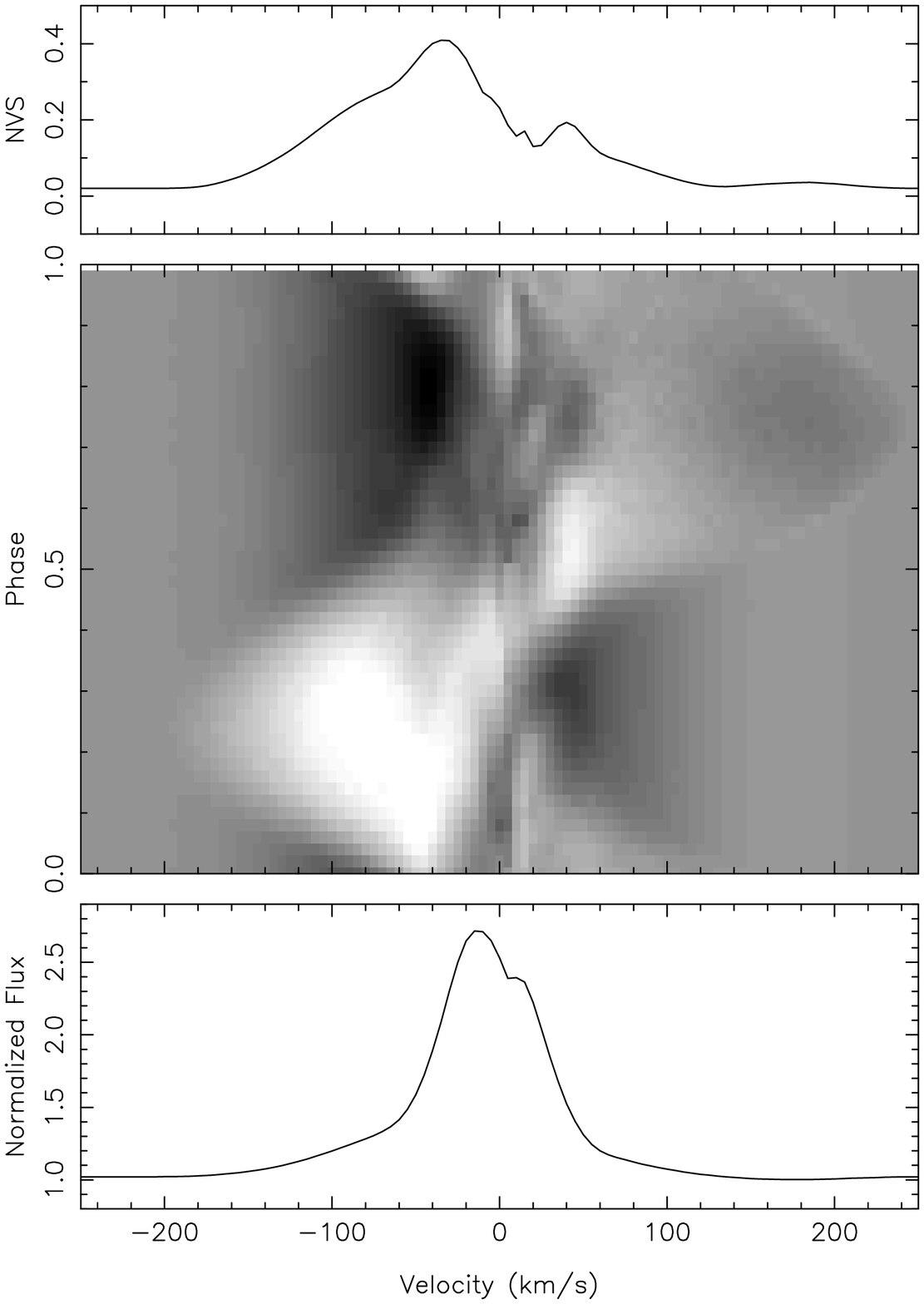}\tabularnewline
\end{tabular}\par\end{center}

\caption{The summary of the Pa$\beta$ spectra computed for Model~A (left),
Model~B (centre) and Model~C (right) which are viewed with inclination
angles ($i$) of $10^{\circ}$, $60^{\circ}$ and $80^{\circ}$ respectively
(c.f.~Fig.~\ref{fig:summary_models_ABC}). All three models have
physically same accretion stream, i.e.~$\Theta=15^{\circ}$. For
each model, spectra were computed at 50 different rotational phases.
In the bottom panels, the mean spectra of all rotational phases are
shown. In the middle panels, the quotient spectra (each spectrum divided
by the mean spectrum) are shown as greyscale images with increasing
rotational phases in upward vertical direction. The greyscale image
is scaled from $1.1$ (white) to $0.9$ (black). 
The normalized variance variance spectra NVS \citep{johns:1995a} 
are shown in the top panels. The
mean spectra for $i=10^{\circ}$ and $60^{\circ}$ are almost symmetric
about the line centre, but they do exhibit a very weak absorption
in their red wings. The greyscale images shows that the red absorption
are sometime stronger (black regions) and some time weaker (white
regions) than the ones seen in the mean spectra. A little absorption
in the red wing is seen for $i=80^{\circ}$ case. The ratio of the
amount of the variability in red wing to that in blue wing decrease
as $i$ increases. }

\label{fig:var_theta_fixed}
				   
\end{figure*} 

\subsection{Dependency on inclination $i$}
\label{sub:Dependency-on-inclination}

Next we examine the dependency of the line variability on inclination
angles using the MHD models with $\Theta=15^{\circ}$.
Fig.~\ref{fig:summary_models_ABC} summarises the
emission maps and profiles (Pa$\beta$) for three different inclination
angles $i=10^{\circ}$, $60^{\circ}$, and $80^{\circ}$  (Models A,
B, and C in Table~\ref{tab:model_parameters} respectively), computed
at four different rotational phases $t=0,0.25,0.5$ and $1.0$.  The
accretion onto the photosphere occurs along two streams (R04);  creating two
hot spots on the surface -- one on each hemispheres. The hot spots are
located about $30^{\circ}$ from the rotational axis and have an
{}`elongated kidney-bean' shape (see also Fig.~2 in R04).  For small inclination cases
(e.g.~$i=10^{\circ}$, Model~A), one spot is clearly visible at all rotational
phases, but for higher inclinations, no spot is visible at certain
rotational phases. The visibility of spots is very important for
formation of the IPC profile and the variability of its absorption
component; hence, the sizes and the location distributions of spots
should be understood for a given model.

The good visibility of the hot spot for $i=10^{\circ}$ (Model~A) results in the
presence of the weak absorption in the red wing of  the model at all
rotational phases; however, the weakness of the absorption is caused
by the unfavourable alignment of spot-funnel-observer line of
sight. For $i=60^{\circ}$ (Model~B) and $80^{\circ}$ (Model~C), the absorption
feature in the red wing becomes most visible at the rotational phase
when a spot is facing towards (i.e.~$t=0.75$) the observer. The largest
amount of red wing absorption occurs at the rotational phase at which
a hot spot is facing the observer and when the spot-funnel-observer
alignment is favourable e.g. for $t=0.75$ and $\Theta=60^{\circ}$.

The intensity maps show that most of the Pa$\beta$ line flux
contribution is from the gas in the funnels just above the hot spots
(displayed in purple). For a high inclination case (e.g.~$i=80^{\circ}$),
the presence of an accretion disc greatly affects the line
profiles. For such cases, the line of sight to the accretion funnel
located below the equatorial plane is obscured by the disc at all rotational
phases. The blue-asymmetry (c.f.~\citealt{folha:2001}) of the profile becomes largest at
$t\sim0.25$ (for $i=80^{\circ}$, Model~C) as little gas is moving toward
an observer at this phase. On contrary, a half rotation later
($t\sim0.75$), the profile does not become as asymmetric as the one at
$t\sim0.25$ since the self-absorption of the photons in the stream
significantly reduces the flux in the red wing.

\begin{figure}

\begin{center}

\includegraphics[clip,width=0.45\textwidth]{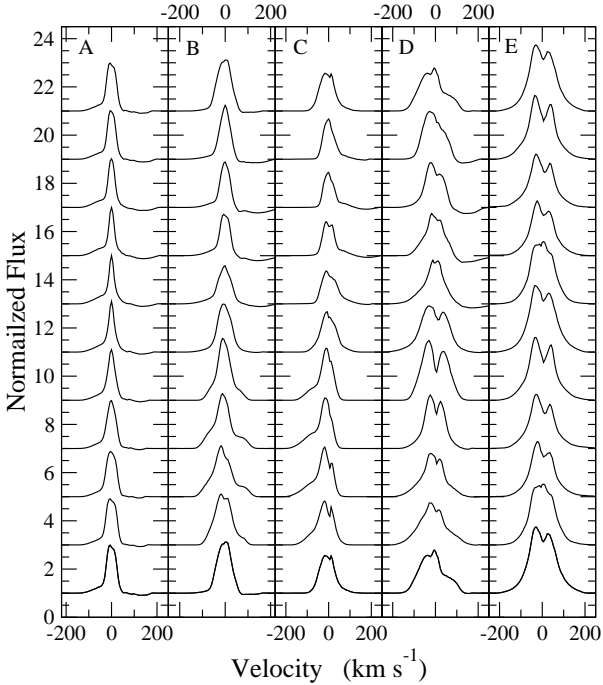}

\end{center}

\caption{The time-series spectra of Pa$\beta$ from Models A through E
(from left to right,) are shown for 10 different rotational
  phase. From the top to the bottom the phase changes from $0$ to $1$. 
Each profile is separated by the rotational phase of
$\sim0.1$, and shifted upward by $1.0$ as the rotational phase
increases for clarity. }

\label{fig:line_var_PaB_all}

\end{figure}

The line variability behaviour of Models~A, B and C are summarised in
Fig.~\ref{fig:var_theta_fixed}. The figure shows the mean spectra
(phase averaged spectra), the quotient spectra (which are the original
profiles divided by the mean spectra) as a function of rotational phase (in 
greyscale image), and 
the normalized variance spectrum (NVS), which is
similar to the root-mean-square spectra
(c.f.~\citealt{johns:1995a}), for each
model. The mean spectra of the 
three models are fairly symmetric about the line centre; however, a
very weak but noticeable amount of absorption in the red wings can be
seen in the spectra at all $i$.  For $i=10^{\circ}$ and $60^{\circ}$
cases (Models~A and B), the flux level in their red wing becomes slightly below the
continuum level, but the level remains above the continuum for the 
$i=80^{\circ}$ case (Model~C). Although the line equivalent width of
the mean spectra for $i=10^{\circ}$ is slightly smaller than that of
$i=60^{\circ}$ and $80^{\circ}$ cases, no significant difference is
seen in three models. In addition to the quotient spectra in
Fig.~\ref{fig:var_theta_fixed}, the phase dependent spectra of each
model are also shown in Fig.~\ref{fig:line_var_PaB_all} as a different
representation of the line variability.

The amount of the flux variability as a function of wavelength is
summarised as the NVS. For $i=10^{\circ}$ 
(Model~A), similar levels of variations are seen in the red and
blue sides.  The NVS is double-peaked, and is symmetric around the
line centre.  This variability pattern (NVS) is very similar to that
of H$\alpha$ and H$\beta$ from the CTTS CW Hydra (K7Ve) presented by
\citet{Alencar:2002}.  The system has a low inclination angle
($i=18^{\circ}\pm10^{\circ}$, \citealt{Alencar:2002}) which is
consistent with our Model~A ($i=10^{\circ}$). The variability
patterns of the red and blue sides are also very similar to each other
as seen in the grey scale image.  As $i$ increases, the symmetry
breaks; the fraction of peak strength on the red to that on the blue
becomes smaller. The peak levels the NVS (on the red side) for
$i=60^{\circ}$ and $80^{\circ}$ are similar ($\sim10$ per~cent), but
are about twice the size in the $i=10^{\circ}$ case.

\subsection{Dependency on misaligned angle $\Theta$}

\label{sub:Dependency-on-misaligned}

Next, we compare the models with three different misalignment angles:
$\Theta=15^{\circ}$, $60^{\circ}$ and $90^{\circ}$ (Models B, D and E
respectively). The inclination angle $i$ is fixed at $60^{\circ}$ for
these three models. Pa$\beta$ profiles at rotational phase $t=0$,
$0.25$, $0.5$ and $0.75$ along with the corresponding spatial
intensity maps are shown in Figs.~\ref{fig:summary_models_ABC} (for
Model~B) and \ref{fig:summary_models_DE} (for Models D and
E). Although the shapes of the funnel flows are different, the
accretion still occurs in two streams for all these models, causing
two hot spots on the stellar surface. The width of the stream becomes
wider as $\Theta$ increases, and consequently the azimuthal extent of
the hot spots also becomes wider. The latitudinal position of the hot
spots becomes lower as the misalignment angle increases
(c.f.~Fig.~\ref{fig:density-zx}). For $\Theta=90^{\circ}$ case
(Model~E), the two wide and thin funnels are located almost on
equatorial plane (c.f.~Fig.~\ref{fig:density-zx}), and so are the hot
spots.  See R04 for larger and clearer depiction of the hot spot
geometries and their physical properties. Interestingly, similar
equatorial flows and the shape of hot spots are found in a quadrupole
magnetic field accretion model of \citet{long:2007} with a small
misalignment angle.

The IPC profiles are found at $t=0.75$ for $\Theta=15^{\circ}$ (Model~B)
and $\Theta=60^{\circ}$ (Model D), but not for $\Theta=90^{\circ}$
(Model~E). The nearly equatorial accretion flows in Model~D do
not have a favourable spot-stream-observer line of sight which is
essential for the formation for of the IPC profile. The double-peaked
profiles are seen for larger $\Theta$ models (Models~D and E). The
splitting of the peaks are caused by the rotational motion of the
magnetosphere.

Figure~\ref{fig:var_i_fixed} shows the summary of the line variability
(Pa$\beta$) for Models B, D and E. The mean line profile becomes
wider as $\Theta$ increases. The separation of two peaks in the mean
profile is largest for $\Theta=90^{\circ}$.
The peak levels of the NVS are similar for all three models, but
the location of the peak(s) is different. While the amount of variability
is largest at the line centre for Model~E ($\Theta=90^{\circ}$),
it is largest in the blue wing ($v\sim40\,\kmps$) for Models~B and
D. We note that variability pattern seen in Model~E resembles that
of H$\beta$ from the CTTS AA~Tau observed by \citet{bouvier:2007} although
their inclination angle ($i\approx 75^{\circ}$) is different
from the model presented here ($i=60^{\circ}$).  See also
Fig.~\ref{fig:line_var_PaB_all} for the summary of the phase dependent 
spectra.

\subsection{Comparison of different lines}

\label{sub:Comparison-of-different}

Once again using the models with $\Theta=15^{\circ}$ (Models~A, B and
C) as examples, we demonstrate the difference in the line shapes among
different hydrogen lines: Pa$\beta$, Br$\gamma$ and H$\beta$.  The
summary of the comparison spectra is shown in
Figure~\ref{fig:summary_diff_lines}. Although they differ in strength,
the overall dependency of the line profile shape on the inclination angle
is very similar in these three transitions.  The relative line
strength slightly decreases from H$\beta$ to Pa$\beta$ then to
Br$\gamma$. The line cores are narrower for higher inclination angle
models. The strength of the blue wings are similar for $i=10^{\circ}$
and $60^{\circ}$, but much weaker in the higher inclination model
$i=80^{\circ}$ since most of the flow moving toward an observer
(blueshifted flows) is eclipsed by the stellar surface at a higher
inclination (c.f.~Fig.~\ref{fig:summary_models_ABC}).  

The line variability behaviour for each line computed at
$i=60^{\circ}$ is summarised in Figure~\ref{fig:var_lines_model_B}.  
All three lines show a very similar variability pattern in both the NVS
spectra and the grey scale quotient spectral images. Interestingly, the
peak level of NVS for H$\beta$ is about 4 
times larger than those of Pa$\beta$ and Br$\gamma$. The difference in
the size of variablity can be explained by a much larger spatial
extent of the line emission regions for H$\beta$ compared to Pa$\beta$
and Br$\gamma$.

\subsection{Line equivalent width}

\label{sub:Line-equivalent-width}

The phase dependency of the line EW (Pa$\beta$) from each model in
Table~\ref{tab:model_parameters} is shown in Figure~\ref{fig:ew_paB}.
We compute three different types of EWs: (1)~by using all 
velocity bins in the model profile (from $-500$ to $500\,\kmps$),
(2)~by using only negative velocity bins (the blue wing: $-500$
to $0\,\kmps$), and (3)~by using only positive velocity bins (the red
wing: $0$ to $500\,\kmps$). Similar methods are often used in
 time-series spectra observations (e.g.~\citealt{johns:1995};
\citealt{kurosawa:2005}) to examine the gas kinematics of the flow. Note
that here we use the convention of the sign for the line EWs as
negative when the flux is below the continuum and positive for vice versa.

For the model with $\Theta=15^{\circ}$ and a mid to low inclination
angle (i.e.~Models A and B), the EW curves show only one
minima (at phase $t=0.75$) in one rotational
period. On the other hand, with the same physical model
($\Theta=15^{\circ}$) but with a high inclination angle
(i.e.~$i=80^{\circ}$ as in Model~C), the shapes of the EWs are
affected by the presence of a second local minima
at $t=0.25$, especially in the red wing (positive velocity bins) EW
curve which clearly shows two local minima in one
rotational period. The local minima at $t=0.75$ is
caused by the maximum continuum flux 
(c.f.~Fig.~\ref{fig:cont_var}) contribution from the hot spot in the
upper hemisphere (c.f.~Fig.~\ref{fig:summary_models_ABC}). The second
local minima at $t=0.25$ is caused by
the combination of the following two reasons. First, at the high
inclination,  only one accretion arm is visible to an
observer. Furthermore, at this rotational phase, the flow component
which is moving away from the observer almost completely disappears
because of the orientation of the remaining upper accretion arm
(c.f.~Fig.~\ref{fig:summary_models_ABC}) hence causing the minimum
flux in the red wing, which corresponds to the minimum EW
of the red wing. Secondly, the visibility of the second hot spot
(c.f.~Section~\ref{sub:Continuum-Variability}) from the lower
hemisphere at this rotational phase causes a slight increase in the
continuum flux as also seen in the light curve in
Fig.~\ref{fig:cont_var}. This slightly weakens the line strength even further.

\begin{figure*}

\begin{center}\begin{tabular}{ccc}
\includegraphics[scale=0.3]{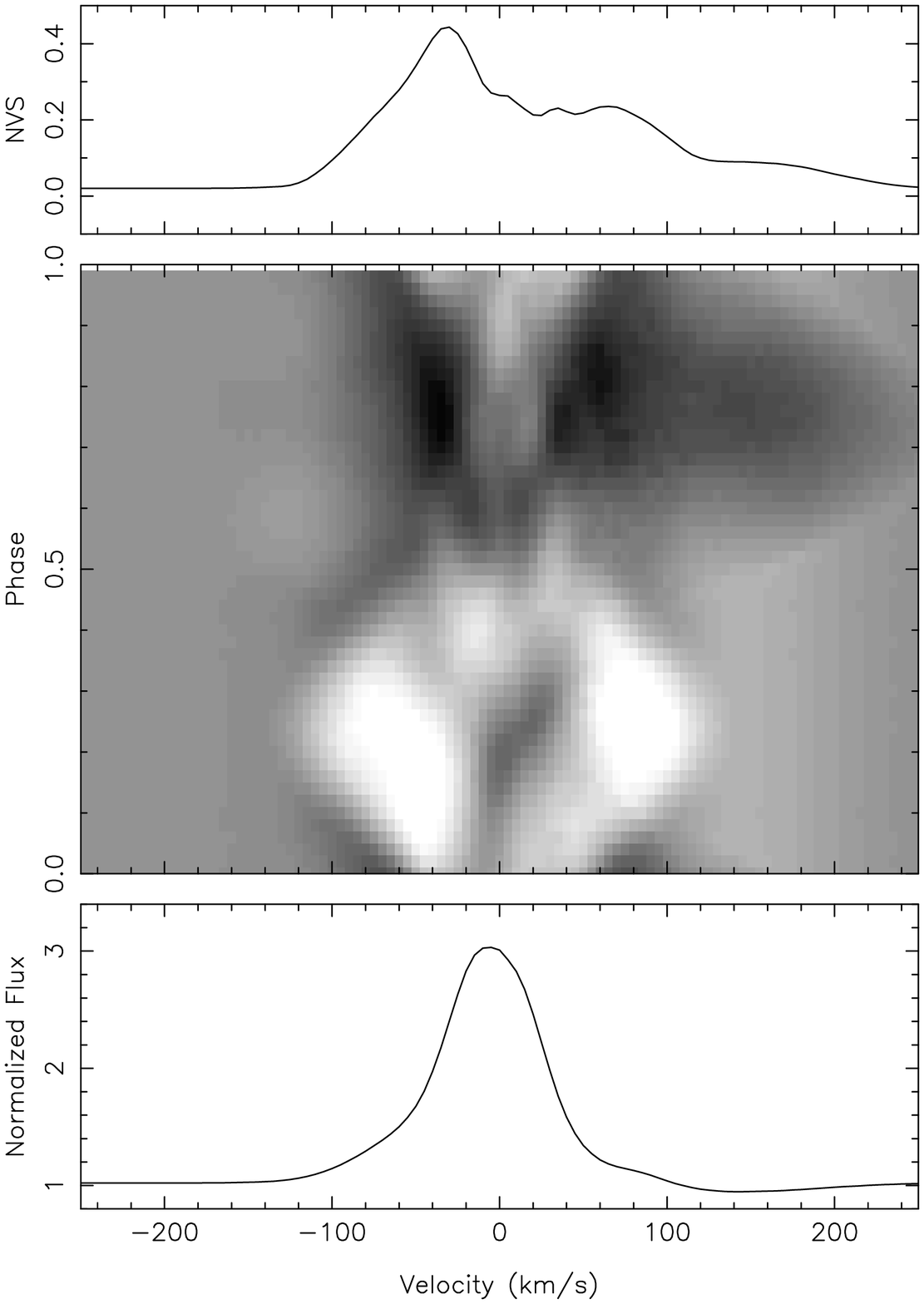}&
\includegraphics[scale=0.3]{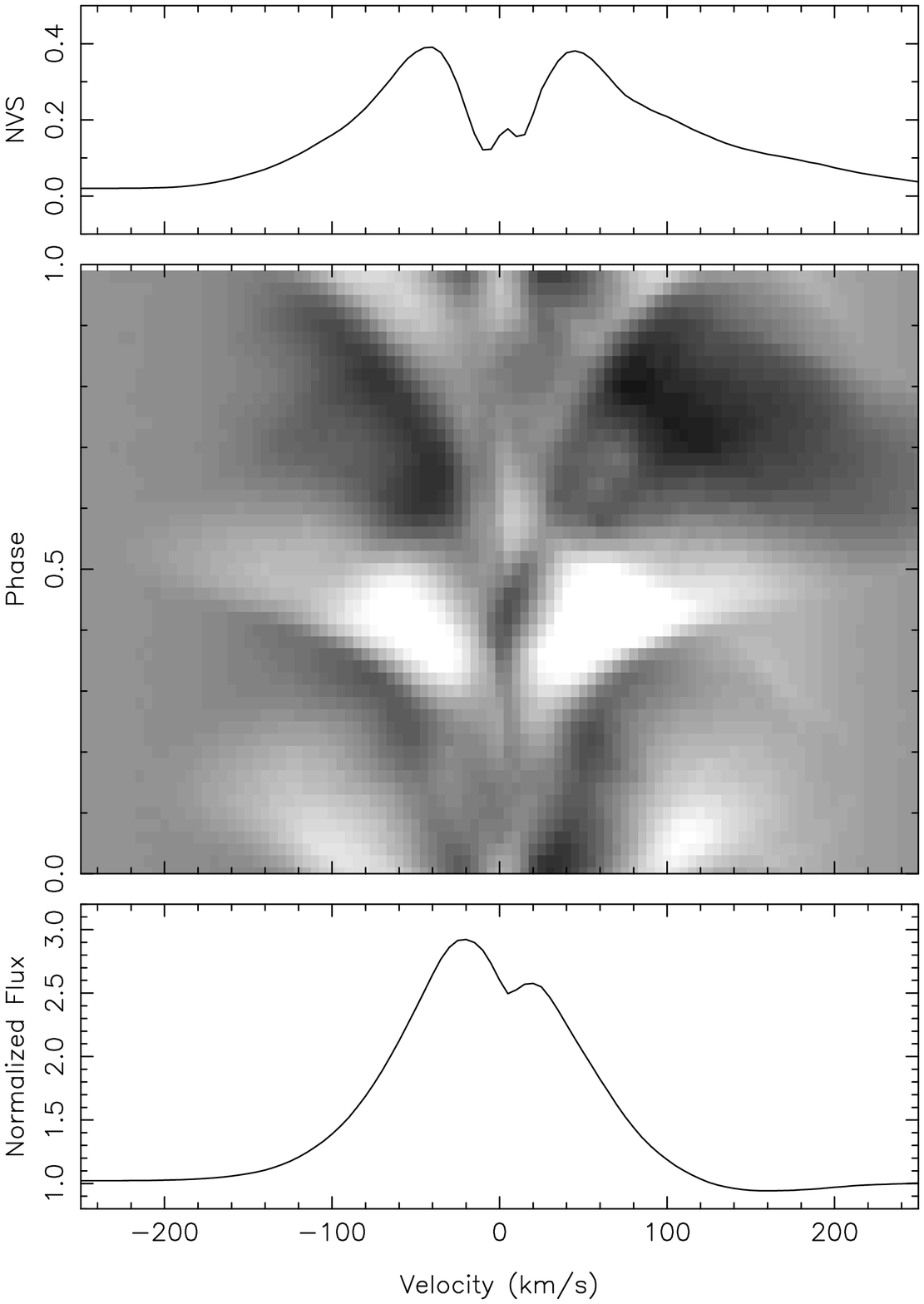}&
\includegraphics[scale=0.3]{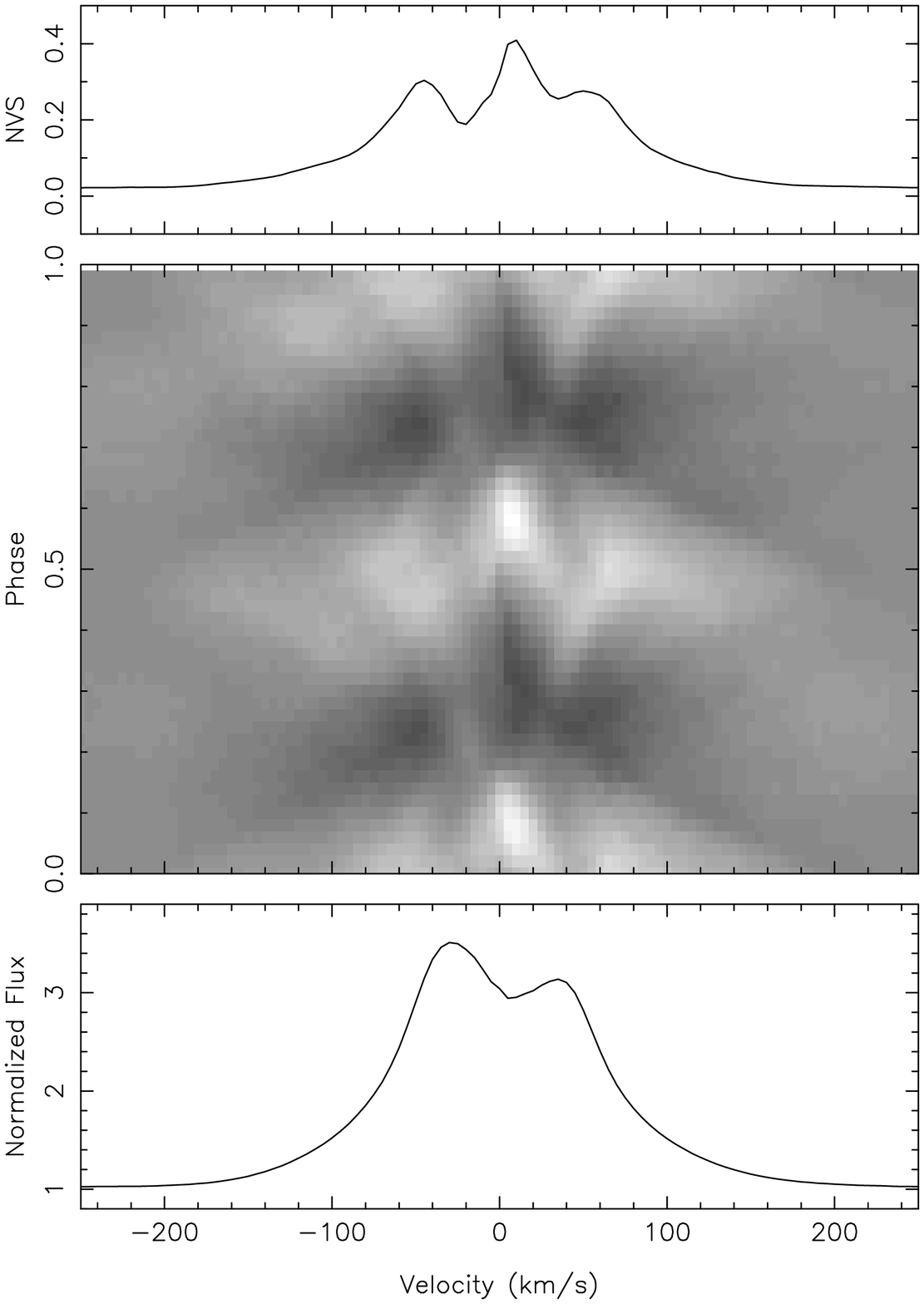}\tabularnewline
\end{tabular}\par\end{center}

\caption{Same as in Fig.~\ref{fig:var_theta_fixed}, but for different magnetic
misalignment angles: $\Theta=15^{\circ}$ (left, Model~B), $60^{\circ}$
(centre, Model~D), and $90^{\circ}$ (right, Model~E). The inclination
is fixed at $i=60^{\circ}$ for these models. The mean spectrum (bottom
panels) becomes wider and more asymmetric (around the line centres)
as $\Theta$ increases. A weak absorption in the red wing are seen
in the mean spectra for $\Theta=15^{\circ}$ and $60^{\circ}$ cases,
but no clear red absorption is seen in $\Theta=90^{\circ}$ case.
The variability patterns (middle and top panels) for three cases
are quite distinct from one another. }

\label{fig:var_i_fixed}
\end{figure*}

\begin{figure}

\begin{center}
    \includegraphics[clip,width=0.45\textwidth]{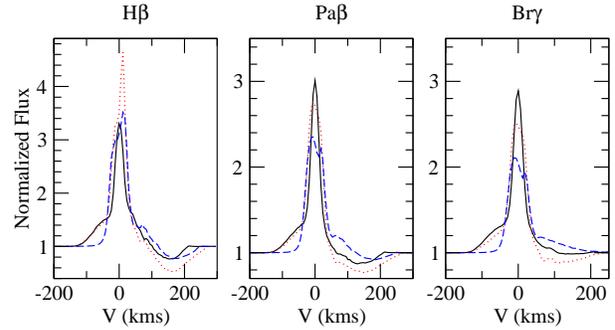}\par
\end{center}

\caption{Comparison of H$\beta$, Pa$\beta$ and Br$\gamma$ for
$i=10^{\circ}$ (Model~A: solid), $60^{\circ}$ (Model~B: dotted) and
$80^{\circ}$ (Model~C: dashed) from left to right respectively. All
the profiles are computed at the rotational phase of 0.75
(c.f. Fig.~\ref{fig:summary_models_ABC}).  Overall dependency on the
inclination angles are similar for all three transitions although the
relative strength slightly decreases from H$\beta$ to Pa$\beta$ then
to Br$\gamma$. The line cores are narrower for higher inclination
angel models. The strength of the blue wings are similar for
$i=10^{\circ}$ and $60^{\circ}$, but much weaker in the higher
inclination model $i=80^{\circ}$ since the most of the flow moving
toward an observer (blueshifted flows) is eclipsed by the stellar
surface in a higher inclination
(c.f.~Fig.~\ref{fig:summary_models_ABC}).  The depth of the red wing
absorption is largest for $i=60^{\circ}$ cases where the aligment of
a hot spot--infall stream-- observer is near optimum.}
\label{fig:summary_diff_lines}

\end{figure}

For larger misalignment models (Models~D and E), we observe two
local minima
in one rotational period. As mentioned earlier, the two
local minima are seen
because of the visibility of two hot spots in these
models at this inclination ($i=60^{\circ}$). Remember
that the continuum light curves of these models
(Fig.~\ref{fig:cont_var}) also show two local minima in one rotational
phase. We note that the position of the local minima in the EW 
for these two models coincide well with the position of 
 the maxima (or local maxima) of the continuum light curves in
Fig.~\ref{fig:cont_var}. This clearly demonstrates that the hot spot
visibility greatly influences the line EW variability.  
To quantify the statement above, we have computed the Pearson's correlation
coefficients ($r$, c.f.~\citealt{bevington:1969}) for the continuum light curves in
Fig.~\ref{fig:cont_var} and the EW curves in Fig.~\ref{fig:ew_paB} for
each model. The results are $r=$ 0.67, 0.92, 0.95, 0.88, 0.99 for
Models~A, B, C, D and E respectively --- indicating fairly good
correlations between the continuum and the EW variations.

The EW of the red wing and that of the blue wing of each model are 
well correlated with each other except for Model~C which has
a very high inclination ($i=80^{\circ}$). The figure shows that during
 half of the rotational period (between $t=0$ and $0.5$), the EW
curves are anti-correlated each other. This anti-correlation is
naturally caused by the fact that only one accretion arm is visible to
an observer at this high inclination. When the arm is on the opposite
side of the star (when the star is in between the observer and the
arm), the EW of the blue wing should be maximum and
that of the red wing should be minimum. When the arm
is on near side of the observer, the opposite 
should occur if the hot spots are not visible to the observer.
As explained earlier, the high visibility of the hot spot on the upper
hemisphere causes the increase in the continuum flux resulting in
lower EW in the red wing for the phase between $t=0.5$ and $1.0$,
causing the correlation between the EW of the red wing and the blue
wing (instead of an anti-correlation).

In the time-series spectra observation of the CTTS SU~Aur,
\cite{johns:1995} found an anti-correlation between the blue wing
EW of H$\alpha$ and the red wing EW of H$\beta$.  They discussed that
the anti-correlation might be explained by a tilted dipole magnetosphere
with outflow components from the disc-magnetosphere interaction
region. Although there is no wind/outflow component in our model, the
mechanism producing the anti-correlation seen in Model~C may be
also an additional reason for the observed anti-correlation in SU~Aur.
We also note that the high inclination which is required for this
mechanism is consistent with the high inclination angle of SU~Aur
(e.g. \citealt{muzerolle:2003}; \citealt{unruh:2004};
\citealt{kurosawa:2005}). 
In  time-series spectroscopic observation of Pa$\beta$ from SU~Aur,
\cite{kurosawa:2005} also found the EW of the red wing
anti-correlates with that of the blue wing at some rotational phases,
but they correlate in other phases. Once again this behaviour is quite
similar to that one seen our Model~C.

\begin{figure*}

\begin{center}\begin{tabular}{ccc}
\includegraphics[clip,scale=0.3]{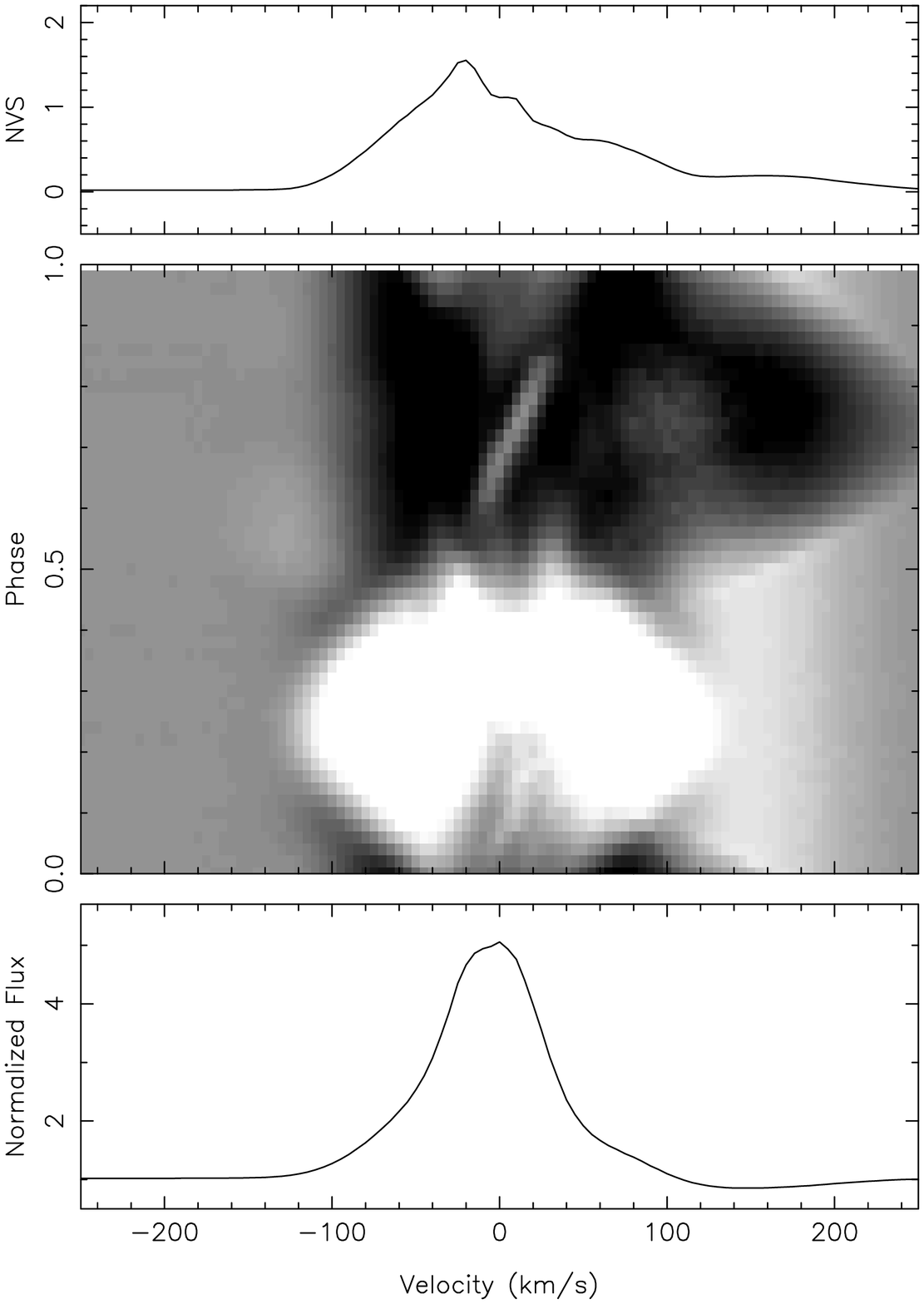}&
\includegraphics[clip,scale=0.3]{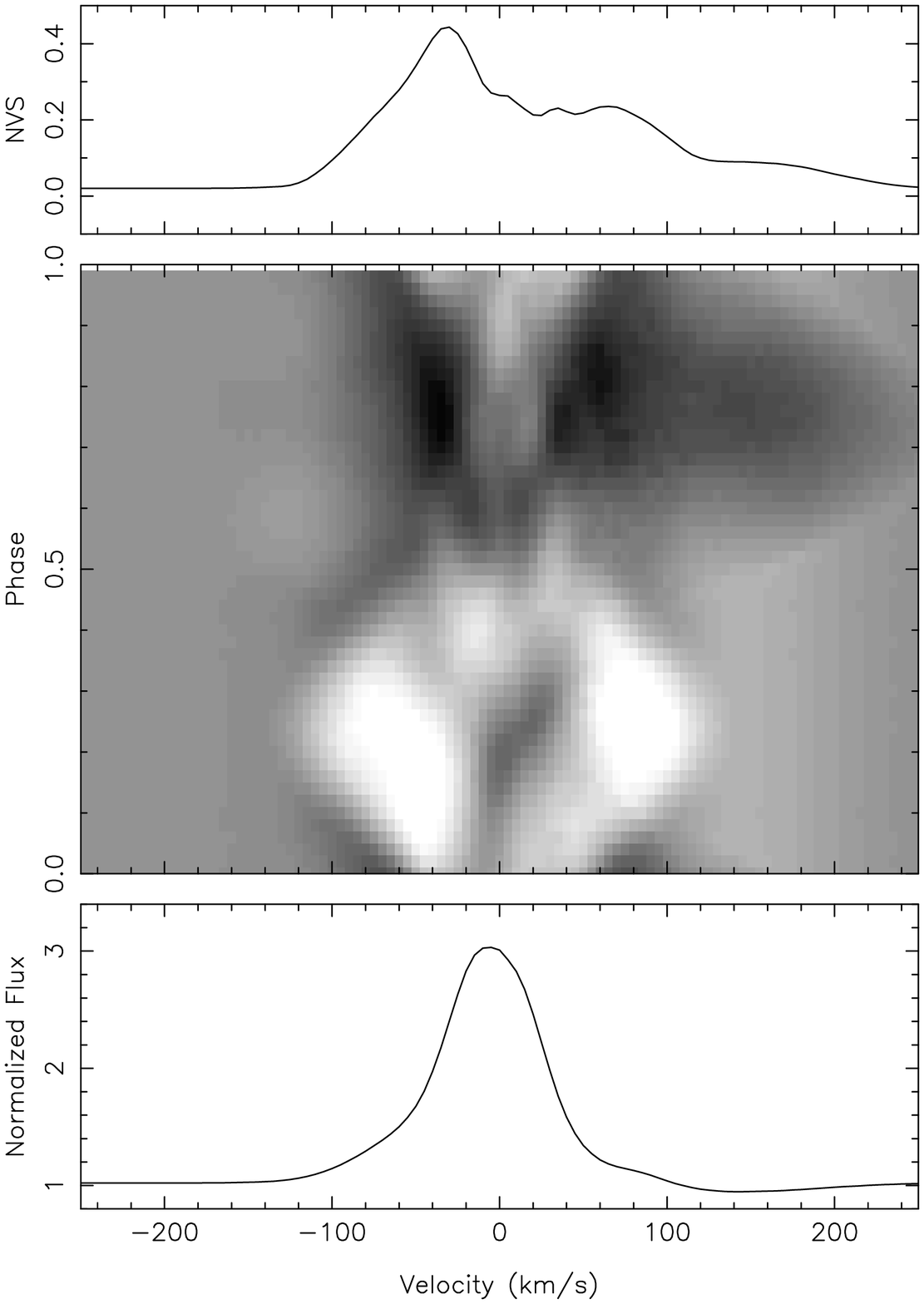}&
\includegraphics[clip,scale=0.3]{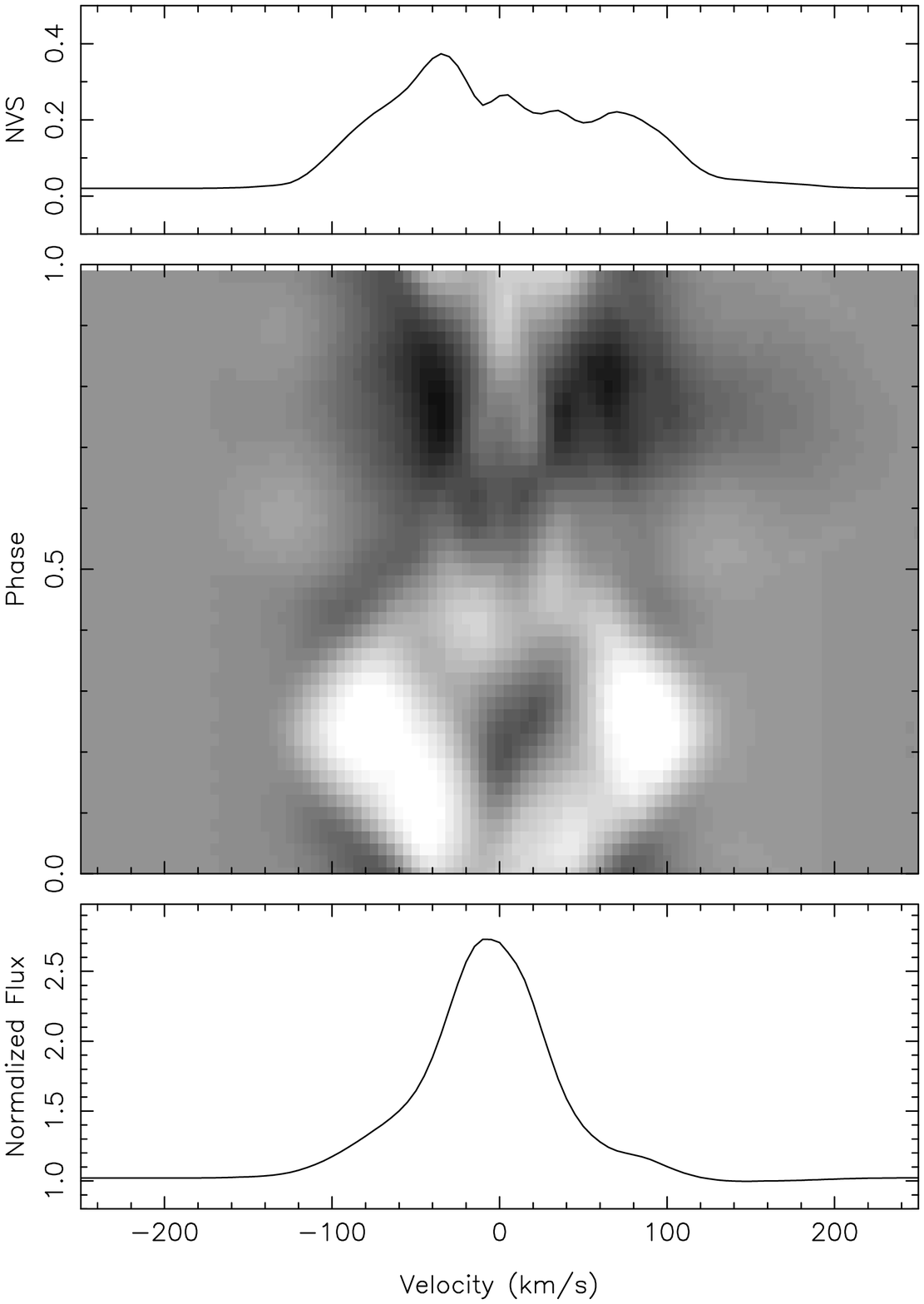}\tabularnewline
\end{tabular}\par\end{center}

\caption{Same as in Fig.~\ref{fig:var_theta_fixed}, but for Models~B
($i=60^{\circ}$ and $\Theta=15^{\circ}$) with different lines:
H$\beta$ (left), Pa$\beta$ (centre) and Br$\gamma$ (right). The line
strength of the mean spectra (bottom panels) decreases from H$\beta$
to Pa$\beta$, and then to Br$\gamma$. All lines show a similar
variability pattern as seen in their quotient spectra greyscale images
(middle panels); however, they differ in their
magnitudes (top panels). }

\label{fig:var_lines_model_B}

\end{figure*}

\section{Discussion}
\label{sec:discussion}

\subsection{Comparison with observations}
\label{sub:compariosn-with-observation}
Although not based on simultaneous observations, \citet{folha:2001}
presented 42 Pa$\beta$ and 30 Br$\gamma$ profiles of CTTS and weak
T~Tauri stars (WTTS) mainly in the Taurus-Auriga complex. 
They found that 53~per~cent of Pa$\beta$ profiles have shapes
symmetric around the line centre (Type~I profile:
\citealt{reipurth:1996}) and 34~per~cent of them have IPC
profiles.  Similarly for Br$\gamma$, they found 72 and 20~per~cent for
Type~I and IPC profiles respectively.  
They also found no blueshifted absorption components in both Pa$\beta$
and Br$\gamma$ except in one object (CW~Tau) and only in Pa$\beta$. 

All of our models in Figs.~\ref{fig:summary_models_ABC} and
\ref{fig:summary_models_DE} (except for Model~E) show the IPC at some
rotational phases. For Br$\gamma$, in general we found that the IPC
feature (the redshifted absorption component) is much weaker than
that seen in the Pa$\beta$. For example,
Fig.~\ref{fig:summary_diff_lines} shows the flux level in the red
wing remains above the continuum for mid to low inclination systems, and
it is below the continuum level only for a very high inclination
system.  Using the time-series model spectra (which are equally
sampled in rotational phase) shown in
Figs.~\ref{fig:var_theta_fixed}, \ref{fig:var_i_fixed} and
\ref{fig:var_lines_model_B}, we have computed the fraction of the
occurrence of IPC profiles in each model. The results are
summarised in Table~\ref{tab:compare} along with the fraction of the
IPC profiles in the Pa$\beta$ and Br$\gamma$ from the observations of
\citet{folha:2001}.  For a fixed $\Theta$ value (Models~A, B and C),
the fraction of the IPC profiles decreases as the inclination angle $i$
increases for Pa$\beta$. The largest inclination model
($i=80^{\circ}$, Model~C) shows the fraction of IPC profiles is 
37~per~cent which is closest to the observed value of
34~per~cent.  On the other hand, no such clear trend is seen for the
Pa$\beta$ models with the fixed $i=60^{\circ}$ value but for different
$\Theta$ values (i.e.~Models~B, D and E).  At this mid-inclination
angle $i=60^{\circ}$ ($\cos i=0.5$), the fraction of IPC profiles (for
Pa$\beta$) is 50, 50 and 0~per~cent for Models B, D and E
respectively.  The fractions for Models~B and D are much larger than
that of the observations, and that for Model~E is much smaller.  To
compare the models more strictly with the IPC fractions of the
observations, one needs enough sampling of the model profiles in
both rotational phase and inclination angle. Although we have
enough phase sampling points (50 angles), we lack a full sampling of  
inclination angle (3 angles); hence, the direct
comparison is difficult here.

The full width half maxima (FWHM) of the mean (phase-averaged)
Pa$\beta$ and Br$\gamma$ spectra of our models (Figs.~\ref{fig:var_theta_fixed},
\ref{fig:var_i_fixed} and \ref{fig:var_lines_model_B}) are also summarised
in Table~\ref{tab:compare} along with the mean  FWHM of the
observed Pa$\beta$ and Br$\gamma$ profiles from \citet{folha:2001}. 
The table clearly shows that in all models presented here,
the predicted FWHM underestimate the observed values.  The discrepancy
may be caused by one or a combination of the following:
(1)~the relatively small size of the magnetosphere 
in the MHD models in which the gas velocities just before reaching the
stellar surface is relatively small $\sim 200~\kmps$, (2)~the spatial widths
of the accretion funnels are underestimated, (3)~other line
broadening mechanisms (besides the Doppler) may be important
(e.g.~\citealt{muzerolle:2001}; \citealt{kurosawa:2006}), and (4)~the
temperature of funnel streams is underestimated.
An implication for possible cause (2) can be seen in the tendency of 
increasing FWHM values as increasing $\Theta$ values. Note that the
widths of the accretion streams become wider for a larger $\Theta$
model.  As mentioned before (Section~\ref{sec:temp_structure}), the
normalisation of the temperature structures used in our models is
quite arbitrary, and the temperatures used here may underestimate
that of a typical CTTS.

We also find that among our models only $\Theta=15^{\circ}$ models (Models~A,
B and C) show rather symmetric Type~I profiles. The larger $\Theta$
models (Models~D and E) show profiles split near the line centres, and
have an appearance of a double-peaked profile. The difference is
mainly caused by the geometry of the accretion funnels and the locations of
the hot spots where the accretion funnels meet the stellar surface.
In Models~D and E with $\Theta=60^{\circ}$ and $90^{\circ}$
respectively, the accretion funnels are much wider and the hot spot
latitudes are much lower compared to those of the $\Theta=15^{\circ}$
models. This would allow an observer to see near the base of the
accretion funnels from both hemisphere for the larger misalignment
angle models (c.f.~Fig.~\ref{fig:summary_models_DE}). This and the
rotational motion of the magnetosphere causes a double-peaked
profiles. On the other hand, in the low misalignment angle models, the
base of only one accretion funnel is visible to an observer
(c.f.~Fig.~\ref{fig:summary_models_ABC}). 
Based on the comparison of the line profile shapes, we find that 
a model with a smaller misalignment angle ($\Theta \sim
15^{\circ}$) is more consistent with the observations of
\citet{folha:2001}. Note that the double-peaked profile shapes
(Type~II-R, c.f. \citealt{reipurth:1996}) seen in Models~D and E are
not entirely inconsistent with the observations, as \citet{folha:2001}
found such Pa$\beta$ profiles in a few systems. 

The hot spot surface coverage fractions in our models (in
Figs.~\ref{fig:summary_models_ABC} and \ref{fig:summary_models_DE})
are approximately 7~per~cent (for Models~A, B and C), 9~per~cent (for
Model~D) and 6~per~cent (for Model~E). These values seem rather large
compared to the hot spot coverage indicated by observations. For
example, \citet{calvet:1998} found that for the majority of CTTS, the
filling factor is $f=0.001-0.01$, and the corresponding surface
coverage of the spots is only 0.1--1~per~cent of the stellar surface
(see also \citealt{gullbring:2000}).  Indeed, these sizes are in
agreement with those derived by \citet*{valenti:1993} using
spectrophotometric observations of `blue continuum.' However,
\citet{muzerolle:1998} pointed out that the filling factor $f$ seems
to be much larger at longer wavelengths, and it was suggested that
possibly the filling factor depends on wavelength.  Here we note that
the hot spots obtained in the 3-D MHD simulations are strongly
inhomogeneous i.e. the kinetic energy per unit area (matter flux) is
much larger in the centre of the spot compared to peripheral regions
of the spot (c.f. see Figs. 2 and 3 in \citealt{romanova:2004}).

In addition, the line strength and shape of the models are not
expected to be very sensitive to a hot spot coverage, provided
that the mass-accretion rate and hence the hot spot luminosity is
unaffected by the change in the size of hot spot
itself. However, if the hot spot coverage becomes much
larger (e.g. 30~per~cent), then the geometry of the
magnetosphere itself must change significantly.  This would
result in a significant change in the line shapes. In general,
the lines are more sensitive to mass-accretion rates and
consequently to hot spot luminosities.  A possible effect of a
reduced hot spot coverage area in the model would be a smaller
chance of producing the IPC profiles. In turn, this would give
us a better agreement with the IPC fraction seen in the
observations of \citet{folha:2001}, as discussed earlier
(c.f.~Table~\ref{tab:compare}).

\begin{figure}

\begin{center}
  \includegraphics[clip,width=0.4\textwidth]{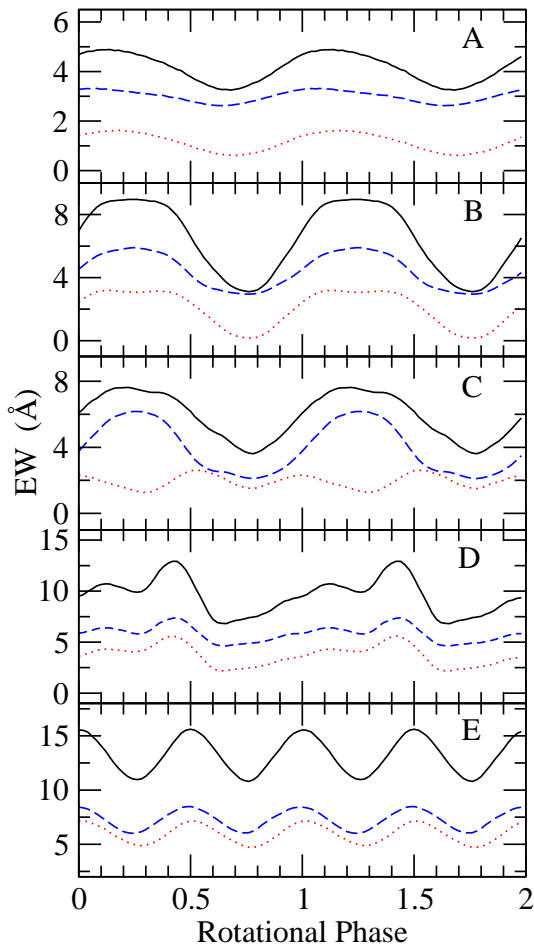}
\end{center}

\caption{Equivalent width (EW) variations as a function of rotational
  phase for the six models
(A through F) in Table~\ref{tab:model_parameters}. For each model,
the total EW (solid), the EW in the red wing (dotted), and that in
the blue wing (dashed) are separately computed. Except for Model~C,
the temporal variations of the EW in the red wing and that in the
blue wing are well correlated. For Model~C, they are anti-correlated
between phase $t=0$ and $0.5$ (see text for explanation).}

\label{fig:ew_paB}

\end{figure}


\begin{table}

\begin{center}
\begin{tabular}{ccc}
\hline 
Model/Obs.&                FWHM&     IPC fraction   \tabularnewline
          &              ($\kmps$)&  (per~cent)     \tabularnewline
\hline 
Observation (Pa$\beta$) &  $204\pm12$&        34\tabularnewline
Model~A (Pa$\beta$)     &  $42$      &       100\tabularnewline
Model~B (Pa$\beta$)     &  $70$      &        50\tabularnewline
Model~C (Pa$\beta$)     &  $70$      &        37\tabularnewline
Model~D (Pa$\beta$)     &  $113$     &        50\tabularnewline
Model~E (Pa$\beta$)     &  $129$     &         0\tabularnewline
Observation (Br$\gamma$)&  $207\pm26$&        20\tabularnewline
Model~B (Br$\gamma$)    &  $70$      &        31\tabularnewline
\hline
\end{tabular}
\end{center}

\caption{Comparison with the observation of \citet{folha:2001}. See
  Table~\ref{tab:model_parameters} for model parameters.}

\label{tab:compare}

\end{table}


\subsection{Intrinsic variability}
\label{sub:intsinsic-variability}

In the previous section (Section~\ref{sec:results}), we presented the
`rotationally modulated' variability of the spectra based on the MHD
simulations. We observed that the variability associated with the
absorption in the red wing reaches up to 30--40~per~cent
(c.f.~Figs.~\ref{fig:var_theta_fixed}, \ref{fig:var_i_fixed} and
\ref{fig:var_lines_model_B}).  Here we briefly examine whether such
rotationally modulated variability can be observed when, in reality,
the accretion rate through the funnels to the stellar surface may also 
vary in time.  The MHD simulations have shown that in many cases the
mass-accretion rate ($\dot M$) to the stellar surface reaches
quasi-stationary states, but the mass flux continues to vary on a
time-scale of 3--5~d, which is similar to the rotationally modulated
time-scale ($\sim 4$~d). Thus the line profile variability induced by
the stellar rotation may be complicated by the intrinsic variation of the
accretion rate. However, the MHD simulations show that in one of the
main cases ($\Theta=15^\circ$ and $\gamma=1.1$), the accretion rate
$\dot M$ is almost constant, and varies only at the level of 
5--6~per~cent. In this case, we expect that the effect of the
intrinsic variability on the line variability is relatively small, and
the results presented in the previous section are not affected 
significantly.  On the other hand, in case for $\Theta=15^\circ$ and
$\gamma=5/3$ (c.f.~\citealt{romanova:2003, romanova:2004}), we find
that the variation of $\dot M$ occurs at the level of
10--30~per~cent.  We expect in this case, the line profile
variability to be affected by the change in $\dot M$, which occurs on
a similar time scale as the rotational period but varies rather
stochastically.  The line variability will be a mixture of periodic
signals from stellar rotation and rather random signals from the
change in $\dot M$.  However, we expect that in this case long
simulation runs will help to separate the contribution from this large
but random component (caused by the change in $\dot M$) from the total
line variability.

\subsection{The Sobolev approximation}
\label{sub:sobolev}

Although not explicitly shown here, the funnel flow velocities near
the inner edge of the accretion disc are rather slow and subsonic.
For example in Models A, B, and C ($\Theta=15^{\circ}$ cases), the
flow velocities are very similar to the one shown in Fig.~5 (right
panel) of \citet{romanova:2004} in which the Mach number of the gas
flow in the middle of the funnel flow is plotted as a function of
distance from the star (along the stream line). The plot shows that
 most of the flow is in fact supersonic except for the gas near
the disc.  The Sobolev Approximation used in the radiative transfer
models is not quite valid in the subsonic part of the flow; however, most of
the line emission occurs in the funnel flow near the stellar surface
(c.f.~Figs.~\ref{fig:summary_models_ABC} and
~\ref{fig:summary_models_DE}), and the line of sight to the emission
region does not pass through the slow moving part of the funnel
flow near the inner edge of accretion disc, except for a very high
inclination case.  For this reason,  the Sobolev approximation used in
the line profile models presented here should be a reasonable
assumption, provided that the intrinsic line width is negligible
compared to the Doppler broadening due to the bulk (macroscopic) motion of gas.

In Section~\ref{sub:compariosn-with-observation}, we noted that the
possibility of additional line broadening mechanisms besides the
the Doppler may be important for explaining much larger FWHMs seen in the
observed Pa$\beta$, compared to those from our models.
\citet{muzerolle:2001} and \citet{kurosawa:2006} considered Stark
broadening which is important in optically thick lines
e.g.~H$\alpha$, but less likely important for Pa$\beta$ and Br$\gamma$
lines. Another possible cause of line broadening is due to large
turbulent motion of plasma which could be caused by  Alfv\'{e}n waves
excited in the magnetosphere (c.f.~\citealt{johns:1995}). If the
magnitude of turbulent velocity is 
comparable to that of the local thermal velocity, the assumption of the
Sobolev approximation becomes invalid.  The MHD simulations
presented in this paper do 
not show such turbulent motions; hence, we have simply adopted the
Sobolev approximation in the work presented here.  On the other hand,
if large turbulence is found in the MHD simulations, the radiative
transfer models should abandon the Sobolev approximation and used an
alternative method e.g.~the ray-by-ray integration method in
\citet{muzerolle:2001} and \citet{kurosawa:2006}.

\subsection{Comparison with earlier models}
\label{sub:earlier-models}

Although very successful in explaining many CTTS emission line
features, the MA model of
\citet{muzerolle:2001} are axi-symmetric (2.5-D); hence, they 
cannot predict the line variability associated with
stellar rotation.  On the other hand, their model still can predict
 line variability caused by a non-constant mass-accretion rate. 
\citet{kurosawa:2006} introduced the disc wind -- MA hybrid
model by combining the models of \citet*{knigge:1995} and \citet{hartmann:1994}
i.e. including both outflows and inflows from and to the
CTTS. The model was successful in reproducing the wide variety of
H$\alpha$ profile shapes seen in observations, and it demonstrated the
importance of a wind or outflow component in determining the profiles
shapes of H$\alpha$.  Their models were also axi-symmetric, and did
not consider line variability.

The same dipole MA model as in
\citet{hartmann:1994} and \citet{muzerolle:2001} are used in
the 3-D Monte Carlo radiative transfer model of \citet{symington:2005}
(see also \citealt{kurosawa:2005}). They have modified the
magnetosphere by removing the flow within some azimuthal angle ranges,
and then displaced the magnetic poles from the rotational poles to
imitate the flow geometry found in the MHD simulation of R03 and R04, to
simulate a possible line 
variability for $i=60^{\circ}$ case. The misalignment angle used in
this model (their A30 model) is rather small ($10^{\circ}$), and it is
very similar to our $\Theta=15^{\circ}$ models.  Although the line strength
is much weaker than our Model~B (which also uses $i=60^{\circ}$), the
variability pattern seen in their RMS spectra and the quotient spectra
image (see their Figure~9) are very similar to ours
(Fig.~\ref{fig:var_theta_fixed}).  The difference in the line strength
is caused by the difference in the adopted mass-accretion rate and the
mass-weighted mean temperature between the two models.

\section{Conclusions}
\label{sec:conclusion}
We have presented a series of 3-D Monte Carlo radiative transfer 
calculations of the line variability from CTTS induced by 
stellar rotation using the results of the 3-D MHD simulations of
\citet{romanova:2003, romanova:2004} who considered the accretion onto
the CTTS with a misaligned dipole axis with respect to the rotational
axis.  Our model here does not include an outflow component, and is 
restricted to the case of accretion flow through the magnetosphere. 
Main objectives here were firstly to examine whether the 3-D MHD simulations
would be able to reproduce the types of line 
profiles and the line variability seen in observations, and secondly to
examine how the line variability depends on basic physical
parameters. In the following, we will summarise our main findings
though this investigation. 

(1)~The detail of the temperature structure along the accretion funnel
   flow does not affect the line shapes greatly
   (Section~\ref{sec:result_temp_structure}). For a fixed
   (density-weighted) mean temperature ($T_{\mathrm{mean}}$) of
   the accretion flow, we have considered three different temperature
   structures (HCH, ACH and isothermal
   c.f.~Section~\ref{sec:temp_structure}), and found the resulting
   Pa$\beta$ profiles are very similar to each other, at least with
   $T_{\mathrm{mean}}=8000\,\Kelvin$. The line shapes are more
   sensitive to the inclination of the systems, the geometry and hence the velocity
   field of the accretion flow. 

(2)~By comparing our models with the atlas of observed Pa$\beta$
   and Br$\gamma$ profiles of \citet{folha:2001}, we found models
   with a smaller misalignment angle (e.g. $\Theta=15^{\circ}$)
   produce rather symmetric profiles which are seen in the majority of
   the observed sample.
   The models with mid to high misalignment angles produces
   double-peaked profiles (Type~II-R: \citealt{reipurth:1996}) which
   are very rare in the sample of \citet{folha:2001}. This may also
   suggest that the majority of the CTTS have a rather small
   misalignment angle.

(3)~We find that the line equivalent width variability is closely related
   with the visibility of the hot spots on the stellar surface
   (Section\ref{sub:Line-equivalent-width}). For a high inclination
   system with a small dipole misalignment angle (e.g.~Model~C), only
   one accretion of funnel (on the upper hemisphere) is visible to
   an observer at any given rotational phase; hence, the
   anti-correlation of the EW in the blue wing and that in the red
   wing can occur during half of a rotational period,
   specifically when the hot spot is not visible to the observer
   (Fig.~\ref{fig:ew_paB}).

(4)~Based on our line profile models, we find that the MHD models of
   \citet{romanova:2003, romanova:2004} are capable of reproducing
   line profile variability behaviour similar to those seen in
   observations (e.g.~\citealt{Alencar:2002}; \citealt{kurosawa:2005};
   \citealt{bouvier:2007}) 
   although the original temperature predicted by the MHD models had
   to be lowered by an arbitrary scaling factor in the radiative transfer
   calculation due to lack of a proper cooling mechanism in the MHD
   model (c.f.~Sections~\ref{sec:temp_structure}).
   
Despite of the relatively good agreement between our models and observed line
variability, several issues still remain.  First, the
assumption of the Sobolev approximation used in the radiative transfer
calculation may not be valid in the funnel flow near the inner edge of
the accretion disc where the speed of flow is substantially subsonic.  Relaxing this
assumption may be more important in the observed flux calculation than
in the source function calculation according to the finding of
\citet{bastian:1980} who compared line profiles computed in 1-D
(spherical geometry), using three different radiative transfer
methods including the Sobolev approximation method.
Detailed investigation of the validity of the Sobolev approximation in
the context of the magnetospheric accretion is underway,
and shall be presented in a future paper (Harries et al., in
preparation). 

Second, the model presented here does not include a wind component.
Although the effect of the wind on the line profile shapes are
expected to be small in Pa$\beta$ and Br$\gamma$, it could potentially
have significant effects on H$\beta$ and H$\alpha$ (e.g.,
\citealt{reipurth:1996}; \citealt{alencar:2000}).  Strong wind
absorption components are also seen the optically thick
\ion{He}{i}~$\lambda$10830 line (\citealt{edwards:2003};
\citealt{dupree:2005}; \citealt{edwards:2006}). These spectroscopic
observations combined with models (e.g., \citealt{matt:2005,
matt:2007}; \citealt*{ferreira:2006}; \citealt{kurosawa:2006};
\citealt*{kwan:2007}) provide us with opportunities to constrain little
known physical properties (e.g.~geometry and temperature) of the
wind/outflow in sub-AU scales.  However, as the MHD models used in this
paper do not include the outflow, we are not be able to explore
the possible effect of the wind contribution to line variability.  In
principle, our radiative transfer model can be applied to a system
which includes both the MA flow and the wind  once a result of
such MHD calculations becomes available.

Strict tests of our models should be performed by quantitative fitting
of the time-series observations of multiple lines (e.g.,
\citealt{johns:1995}; \citealt{unruh:2004}; \citealt{bouvier:2007})
with good rotational phase coverages. The different lines are formed
in different volumes within the funnels and give us additional
information on the flow kinematics. The spectra from different
rotational phases provides information on how the magnetosphere
changes in azimuthal direction.  Once the validity of the model is
established, it can be used to constrain the complex geometry of the
MA flows around CTTS.  It would be very
interesting to compare the geometry constrained by this method to the
ones determined from the field extrapolation method
(e.g.~\citealt*{jardine:2002}; \citealt{gregory:2006}) using the stellar
surface magnetic field information via the Zeeman-Doppler imaging
technique, obtained with a modern stellar spectropolarimeter such as ESPaDOnS
(\citealt{donati:2006, donati:2007}).

While our underlying assumption of the magnetic field geometry is pure
dipole, a more complex field geometry, such as in \cite{long:2007} who
considered the combination of a dipole and a quadrupole fields in
their MHD models, should be also considered in the line variability
calculations. For example, some CTTS are known to posses non-dipole
magnetic fields (c.f.~\citealt{johnskrull:1999};
\citealt{valenti:2004}; \citealt{donati:2007}).
Finally, we shall also consider the case with a non-constant
mass-accretion rate model. For example, a recent MHD calculation by
Kulkarni \& Romanova in preparation (see \citealt{romanova:2007}) has shown that variable
mass-accretion rates could occur due to instabilities (Rayleigh-Taylor
and Kelvin-Helmholtz) in the MA flow. These add
another level of complexity in the interpretation of observed line
profiles, and their variability.

\section*{Acknowledgements} 

We thank the referee, Dr.~Chris Johns-Krull, who provided us with
valuable comments and suggestions which improved the clarity of the
manuscript.  This work is partially supported by PPARC rolling grant
PP/C501609/1, and benefited from a research visit by RK funded by
Exeter's visitors grant PP/D001617/1. This work was also supported in
part by NASA grants NAG5-13220, NAG5-13060, and by NSF grant
AST-0307817, AST-0507760 and AST-0607135.  This work was also
supported by NASA through grant HST-AR-10680 from the Space Telescope
Science Institute, which is operated by the Association of
Universities for Research in Astronomy, Inc., under NASA contract
NAS5-26555.  The numerical computations presented in this paper were
performed partly on the NASA computing facilities, specifically
Explore and Columbia.  RK thanks Prof.~Richard Lovelace for his
hospitality during a visit at Cornell University. RK is grateful for
Dr.~Daniel Proga for helpful discussions and enthusiastic support for
this work, and is also grateful for Dr.~Neil H. Symington for his
effort in the development of crucial parts of the code TORUS.


\begin{thebibliography}{}

\bibitem[\protect\citeauthoryear{{Alencar} \& {Basri}}{{Alencar} \&
  {Basri}}{2000}]{alencar:2000}
{Alencar} S.~H.~P.,  {Basri} G.,  2000, \aj, 119, 1881

\bibitem[\protect\citeauthoryear{Alencar \& Batalha}{Alencar \&
  Batalha}{2002}]{Alencar:2002}
Alencar S.~H.~P.,  Batalha C.,  2002, \apj, 571, 378

\bibitem[\protect\citeauthoryear{{Bastian}, {Bertout}, {Stenholm} \&
  {Wehrse}}{{Bastian} et~al.}{1980}]{bastian:1980}
{Bastian} U.,  {Bertout} C.,  {Stenholm} L.,    {Wehrse} R.,  1980, \aap, 86,
  105

\bibitem[{{Bevington}(1969)}]{bevington:1969}
{Bevington} P.~R., 1969, {Data reduction and error analysis for the
  physical sciences}. McGraw-Hill, New York

\bibitem[\protect\citeauthoryear{{Bouvier}, {Alencar}, {Harries}, {Johns-Krull}
  \& {Romanova}}{{Bouvier} et~al.}{2007a}]{ppv:2007}
{Bouvier} J.,  {Alencar} S.~H.~P.,  {Harries} T.~J.,  {Johns-Krull} C.~M.,
  {Romanova} M.~M.,  2007a, in {Reipurth} B.,  {Jewitt} D.,   {Keil} K.,  eds,
  Protostars and Planets. University of Arizona Press, Tucson, p.~479

\bibitem[\protect\citeauthoryear{{Bouvier} et~al.}{{Bouvier} et~al.}{2007b}]{bouvier:2007}
{Bouvier} J. et~al., 2007b, \aap, 463, 1017

\bibitem[\protect\citeauthoryear{Calvet \& Gullbring}{Calvet \&
  Gullbring}{1998}]{calvet:1998}
Calvet N.,  Gullbring E.,  1998, \apj, 509, 802

\bibitem[\protect\citeauthoryear{{Calvet}, {Muzerolle}, {Brice{\~n}o},
  {Hern{\'a}ndez}, {Hartmann}, {Saucedo} \& {Gordon}}{{Calvet}
  et~al.}{2004}]{calvet:2004}
{Calvet} N.,  {Muzerolle} J.,  {Brice{\~n}o} C.,  {Hern{\'a}ndez} J.,
  {Hartmann} L.,  {Saucedo} J.~L.,    {Gordon} K.~D.,  2004, \aj, 128, 1294

\bibitem[\protect\citeauthoryear{{Camenzind}}{{Camenzind}}{1990}]{camenzind:1990}
{Camenzind} M.,  1990, in {Klare} G.,  ed., Reviews in Modern Astronomy Vol.~3,
  {Magnetized Disk-Winds and the Origin of Bipolar Outflows.}.
Springer, Berlin, p.~234

\bibitem[\protect\citeauthoryear{{Donati}, {Forveille}, {Cameron}, {Barnes},
  {Delfosse}, {Jardine} \& {Valenti}}{{Donati} et~al.}{2006}]{donati:2006}
{Donati} J.-F.,  {Forveille} T.,  {Cameron} A.~C.,  {Barnes} J.~R.,  {Delfosse}
  X.,  {Jardine} M.~M.,    {Valenti} J.~A.,  2006, Science, 311, 633

\bibitem[{{Donati} {et~al.}(2007){Donati}, {Jardine}, {Gregory},
    {Petit}, {Bouvier}, {Dougados}, {M{\'e}nard}, {Cameron}, {Harries},
    {Jeffers}, \& {Paletou}}]{donati:2007} {Donati} et al., 2007, \mnras, 380, 1297

\bibitem[\protect\citeauthoryear{{Dupree}, {Brickhouse}, {Smith} \&
  {Strader}}{{Dupree} et~al.}{2005}]{dupree:2005}
{Dupree} A.~K.,  {Brickhouse} N.~S.,  {Smith} G.~H.,    {Strader} J.,  2005,
  \apjl, 625, L131

\bibitem[\protect\citeauthoryear{{Edwards}, {Fischer}, {Hillenbrand} \&
  {Kwan}}{{Edwards} et~al.}{2006}]{edwards:2006}
{Edwards} S.,  {Fischer} W.,  {Hillenbrand} L.,    {Kwan} J.,  2006, \apj, 646,
  319

\bibitem[\protect\citeauthoryear{{Edwards}, {Fischer}, {Kwan}, {Hillenbrand} \&
  {Dupree}}{{Edwards} et~al.}{2003}]{edwards:2003}
{Edwards} S.,  {Fischer} W.,  {Kwan} J.,  {Hillenbrand} L.,    {Dupree} A.~K.,
  2003, \apjl, 599, L41

\bibitem[\protect\citeauthoryear{Edwards, {Hartigan}, {Ghandour} \&
  {Andrulis}}{Edwards et~al.}{1994}]{edwards:1994}
Edwards S.,  {Hartigan} P.,  {Ghandour} L.,    {Andrulis} C.,  1994, \aj, 108,
  1056

\bibitem[\protect\citeauthoryear{{Ferreira}, {Dougados} \& {Cabrit}}{{Ferreira}
  et~al.}{2006}]{ferreira:2006}
{Ferreira} J.,  {Dougados} C.,    {Cabrit} S.,  2006, \aap, 453, 785

\bibitem[\protect\citeauthoryear{{Folha} \& {Emerson}}{{Folha} \&
  {Emerson}}{2001}]{folha:2001}
{Folha} D.~F.~M.,  {Emerson} J.~P.,  2001, \aap, 365, 90

\bibitem[\protect\citeauthoryear{Fullerton, Gies \& Bolton}{Fullerton
  et~al.}{1996}]{fullerton:1996}
Fullerton A.~W.,  Gies D.~R.,    Bolton C.~T.,  1996, \apjs, 103, 475

\bibitem[\protect\citeauthoryear{{Gregory}, {Jardine}, {Simpson} \&
  {Donati}}{{Gregory} et~al.}{2006}]{gregory:2006}
{Gregory} S.~G.,  {Jardine} M.,  {Simpson} I.,    {Donati} J.-F.,  2006,
  \mnras, 371, 999

\bibitem[\protect\citeauthoryear{Gullbring, {Calvet}, {Muzerolle} \&
  {Hartmann}}{Gullbring et~al.}{2000}]{gullbring:2000}
Gullbring E.,  {Calvet} N.,  {Muzerolle} J.,    {Hartmann} L.,  2000, \apj,
  544, 927

\bibitem[\protect\citeauthoryear{Gullbring, {Hartmann}, {Briceno} \&
  {Calvet}}{Gullbring et~al.}{1998}]{gullbring:1998}
Gullbring E.,  {Hartmann} L.,  {Briceno} C.,    {Calvet} N.,  1998, \apj, 492,
  323

\bibitem[\protect\citeauthoryear{Harries}{Harries}{2000}]{harries:2000}
Harries T.~J.,  2000, MNRAS, 315, 722

\bibitem[\protect\citeauthoryear{Hartmann, Avrett \& Edwards}{Hartmann
  et~al.}{1982}]{hartmann:1982}
Hartmann L.,  Avrett E.,    Edwards S.,  1982, \apj, 261, 279

\bibitem[\protect\citeauthoryear{{Hartmann}, {Calvet}, {Gullbring} \&
  {D'Alessio}}{{Hartmann} et~al.}{1998}]{hartmann:1998}
{Hartmann} L.,  {Calvet} N.,  {Gullbring} E.,    {D'Alessio} P.,  1998, \apj,
  495, 385

\bibitem[\protect\citeauthoryear{Hartmann, Hewett \& Calvet}{Hartmann
  et~al.}{1994}]{hartmann:1994}
Hartmann L.,  Hewett R.,    Calvet N.,  1994, \apj, 426, 669

\bibitem[\protect\citeauthoryear{{Herbst}, {Bailer-Jones}, {Mundt},
  {Meisenheimer} \& {Wackermann}}{{Herbst} et~al.}{2002}]{herbst:2002}
{Herbst} W.,  {Bailer-Jones} C.~A.~L.,  {Mundt} R.,  {Meisenheimer} K.,
  {Wackermann} R.,  2002, \aap, 396, 513

\bibitem[\protect\citeauthoryear{Hillier}{Hillier}{1991}]{hillier:1991}
Hillier D.~J.,  1991, \aap, 247, 455

\bibitem[\protect\citeauthoryear{{Jardine}, {Collier Cameron} \&
  {Donati}}{{Jardine} et~al.}{2002}]{jardine:2002}
{Jardine} M.,  {Collier Cameron} A.,    {Donati} J.-F.,  2002, \mnras, 333, 339


\bibitem[{{Johns} \& {Basri}(1995)}]{johns:1995a}
{Johns} C.~M., {Basri} G., 1995, \aj, 109, 2800


\bibitem[\protect\citeauthoryear{Johns \& Basri}{Johns \&
  Basri}{1995}]{johns:1995}
Johns C.~M.,  Basri G.,  1995, \apj, 449, 341

\bibitem[{{Johns-Krull}(2007)}]{johnskrull:2007}
{Johns-Krull} C.~M., 2007, \apj, 664, 975

\bibitem[\protect\citeauthoryear{{Johns-Krull}, {Valenti}, {Hatzes} \&
  {Kanaan}}{{Johns-Krull} et~al.}{1999}]{johnskrull:1999}
{Johns-Krull} C.~M.,  {Valenti} J.~A.,  {Hatzes} A.~P.,    {Kanaan} A.,  1999,
  \apjl, 510, L41

\bibitem[\protect\citeauthoryear{{Klein} \& {Castor}}{{Klein} \&
  {Castor}}{1978}]{klein:1978}
{Klein} R.~I.,  {Castor} J.~I.,  1978, \apj, 220, 902

\bibitem[\protect\citeauthoryear{{Knigge}, {Woods} \& {Drew}}{{Knigge}
  et~al.}{1995}]{knigge:1995}
{Knigge} C.,  {Woods} J.~A.,    {Drew} J.~E.,  1995, \mnras, 273, 225


\bibitem[\protect\citeauthoryear{{Koenigl}}{{Koenigl}}{1991}]{koenigl:1991}
{Koenigl} A.,  1991, \apjl, 370, L39


\bibitem[\protect\citeauthoryear{{Koldoba}, {Romanova}, {Ustyugova} \&
  {Lovelace}}{{Koldoba} et~al.}{2002}]{koldoba:2002}
{Koldoba} A.~V.,  {Romanova} M.~M.,  {Ustyugova} G.~V.,    {Lovelace} R.~V.~E.,
   2002, \apjl, 576, L53

\bibitem[\protect\citeauthoryear{{Kulkarni} \& {Romanova}}{{Kulkarni} \&
  {Romanova}}{2005}]{kulkarni:2005}
{Kulkarni} A.~K.,  {Romanova} M.~M.,  2005, \apj, 633, 349

\bibitem[\protect\citeauthoryear{{Kurosawa}, {Harries}, {Bate} \&
  {Symington}}{{Kurosawa} et~al.}{2004}]{kurosawa:2004}
{Kurosawa} R.,  {Harries} T.~J.,  {Bate} M.~R.,    {Symington} N.~H.,  2004,
  \mnras, 351, 1134

\bibitem[\protect\citeauthoryear{{Kurosawa}, {Harries} \&
  {Symington}}{{Kurosawa} et~al.}{2005}]{kurosawa:2005}
{Kurosawa} R.,  {Harries} T.~J.,    {Symington} N.~H.,  2005, \mnras, 358, 671

\bibitem[\protect\citeauthoryear{{Kurosawa}, {Harries} \&
  {Symington}}{{Kurosawa} et~al.}{2006}]{kurosawa:2006}
{Kurosawa} R.,  {Harries} T.~J.,    {Symington} N.~H.,  2006, \mnras, 370, 580

\bibitem[\protect\citeauthoryear{{Kurucz}}{{Kurucz}}{1979}]{kurucz:1979}
{Kurucz} R.~L.,  1979, \apjs, 40, 1

\bibitem[\protect\citeauthoryear{{Kwan}, {Edwards} \& {Fischer}}{{Kwan}
  et~al.}{2007}]{kwan:2007}
{Kwan} J.,  {Edwards} S.,    {Fischer} W.,  2007, \apj, 657, 897

\bibitem[{{Long} {et~al.}(2005){Long}, {Romanova}, \&
    {Lovelace}}]{long:2005}
{Long} M., {Romanova} M.~M., {Lovelace} R.~V.~E., 2005, \apj, 634,
    1214

\bibitem[\protect\citeauthoryear{{Long}, {Romanova} \& {Lovelace}}{{Long}
  et~al.}{2007}]{long:2007}
{Long} M.,  {Romanova} M.~M.,    {Lovelace} R.~V.~E.,  2007, \mnras, 374, 436

\bibitem[\protect\citeauthoryear{{Martin}}{{Martin}}{1996}]{martin:1996}
{Martin} S.~C.,  1996, \apj, 470, 537

\bibitem[\protect\citeauthoryear{{Matt} \& {Pudritz}}{{Matt} \&
  {Pudritz}}{2005}]{matt:2005}
{Matt} S.,  {Pudritz} R.~E.,  2005, \apjl, 632, L135

\bibitem[\protect\citeauthoryear{{Matt} \& {Pudritz}}{{Matt} \&
  {Pudritz}}{2007}]{matt:2007}
{Matt} S.,  {Pudritz} R.~E.,  2007, preprint (astro-ph/0707.0306)

\bibitem[\protect\citeauthoryear{{Muzerolle}, {Calvet} \&
  {Hartmann}}{{Muzerolle} et~al.}{1998a}]{muzerolle:1998a}
{Muzerolle} J.,  {Calvet} N.,    {Hartmann} L.,  1998a, \apj, 492, 743

\bibitem[\protect\citeauthoryear{{Muzerolle}, {Calvet} \&
  {Hartmann}}{{Muzerolle} et~al.}{2001}]{muzerolle:2001}
{Muzerolle} J.,  {Calvet} N.,    {Hartmann} L.,  2001, \apj, 550, 944

\bibitem[\protect\citeauthoryear{{Muzerolle}, {Calvet}, {Hartmann} \&
  {D'Alessio}}{{Muzerolle} et~al.}{2003}]{muzerolle:2003}
{Muzerolle} J.,  {Calvet} N.,  {Hartmann} L.,    {D'Alessio} P.,  2003, \apjl,
  597, L149

\bibitem[\protect\citeauthoryear{{Muzerolle}, {Hartmann} \&
  {Calvet}}{{Muzerolle} et~al.}{1998b}]{muzerolle:1998}
{Muzerolle} J.,  {Hartmann} L.,    {Calvet} N.,  1998b, \aj, 116, 455

\bibitem[\protect\citeauthoryear{{Reipurth}, {Pedrosa} \& {Lago}}{{Reipurth}
  et~al.}{1996}]{reipurth:1996}
{Reipurth} B.,  {Pedrosa} A.,    {Lago} M.~T.~V.~T.,  1996, \aaps, 120, 229


\bibitem[{{Romanova} {et~al.}(2007){Romanova}, {Kulkarni}, \&
  {Lovelace}}]{romanova:2007}
{Romanova} M.~M., {Kulkarni} A.~K., {Lovelace} R.~V.~E., 2007 (astro-ph/0711.0418)


\bibitem[\protect\citeauthoryear{{Romanova}, {Ustyugova}, {Koldoba} \&
  {Lovelace}}{{Romanova} et~al.}{2002}]{romanova:2002}
{Romanova} M.~M.,  {Ustyugova} G.~V.,  {Koldoba} A.~V.,    {Lovelace} R.~V.~E.,
   2002, \apj, 578, 420 (R02)

\bibitem[\protect\citeauthoryear{{Romanova}, {Ustyugova}, {Koldoba} \&
  {Lovelace}}{{Romanova} et~al.}{2004}]{romanova:2004}
{Romanova} M.~M.,  {Ustyugova} G.~V.,  {Koldoba} A.~V.,    {Lovelace} R.~V.~E.,
   2004, \apj, 610, 920 (R04)

\bibitem[\protect\citeauthoryear{{Romanova}, {Ustyugova}, {Koldoba}, {Wick} \&
  {Lovelace}}{{Romanova} et~al.}{2003}]{romanova:2003}
{Romanova} M.~M.,  {Ustyugova} G.~V.,  {Koldoba} A.~V.,  {Wick} J.~V.,
  {Lovelace} R.~V.~E.,  2003, \apj, 595, 1009 (R03)

\bibitem[\protect\citeauthoryear{{Rybicki} \& {Hummer}}{{Rybicki} \&
  {Hummer}}{1978}]{rybicki:1978}
{Rybicki} G.~B.,  {Hummer} D.~G.,  1978, \apj, 219, 654

\bibitem[\protect\citeauthoryear{{Symington}, {Harries} \&
  {Kurosawa}}{{Symington} et~al.}{2005a}]{symington:2005}
{Symington} N.~H.,  {Harries} T.~J.,    {Kurosawa} R.,  2005a, \mnras, 356, 1489

\bibitem[\protect\citeauthoryear{{Symington}, {Harries}, {Kurosawa} \&
  {Naylor}}{{Symington} et~al.}{2005b}]{symington:2005a}
{Symington} N.~H.,  {Harries} T.~J.,  {Kurosawa} R.,    {Naylor} T.,  2005b,
  \mnras, 358, 977

\bibitem[\protect\citeauthoryear{{Unruh}}{{Unruh}}{2004}]{unruh:2004}
{Unruh} Y.~C. et~al.,  2004, \mnras, 348, 1301

\bibitem[\protect\citeauthoryear{{Ustyugova}, {Koldoba}, {Romanova},
  {Chechetkin} \& {Lovelace}}{{Ustyugova} et~al.}{1999}]{ustyugova:1999}
{Ustyugova} G.~V.,  {Koldoba} A.~V.,  {Romanova} M.~M.,  {Chechetkin} V.~M.,
  {Lovelace} R.~V.~E.,  1999, \apj, 516, 221


\bibitem[{{Ustyugova} {et~al.}(2006){Ustyugova}, {Koldoba},
    {Romanova}, \& {Lovelace}}]{ustyugova:2006} {Ustyugova} G.~V.,
    {Koldoba} A.~V., {Romanova} M.~M., {Lovelace} 
    R.~V.~E., 2006, \apj, 646, 304


\bibitem[{{Valenti} {et~al.}(1993){Valenti}, {Basri}, \&
  {Johns}}]{valenti:1993}
{Valenti} J.~A., {Basri} G., {Johns} C.~M., 1993, \aj, 106, 2024

\bibitem[{{Valenti} \& {Johns-Krull}(2004)}]{valenti:2004}
{Valenti} J.~A., {Johns-Krull} C.~M., 2004, \apss, 292, 619

\end{thebibliography}

\end{document}